\documentclass[showpacs,preprintnumbers,amsmath,amssymb]{revtex4}
\usepackage{graphicx}
\usepackage{dcolumn}
\usepackage{bm}
\usepackage{braket}
\newcommand{\be}{\begin{eqnarray}}
\newcommand{\ee}{\end{eqnarray}}

\begin{document}
\title{
Edge states and topological phases in non-Hermitian systems
}
\author{Kenta Esaki$^1$, Masatoshi Sato$^1$, Kazuki Hasebe$^2$, 
and Mahito Kohmoto$^1$}

\affiliation{$^1$Institute for Solid State Physics, University of Tokyo, Kashiwanoha 5-1-5, Kashiwa, Chiba 277-8581, Japan \\
$^2$Department of General Education, Kagawa National College of Technology, 
 Mitoyo, Kagawa 769-1192, Japan
}%
\begin{abstract}

Topological stability of the edge states is investigated for
non-Hermitian systems.
We examine two classes of non-Hermitian Hamiltonians
supporting real bulk eigenenergies in weak non-Hermiticity: 
$SU(1,1)$ and $SO(3,2)$ Hamiltonians.
As an $SU(1,1)$ Hamiltonian, the tight-binding model on
 the honeycomb lattice with
 imaginary on-site potentials 
is examined. Edge states with
Re$E=0$ and their topological stability are discussed by the
winding number and the index theorem, based on the
 pseudo-anti-Hermiticity of the system.
As a higher symmetric generalization of $SU(1,1)$ Hamiltonians,
we also consider $SO(3,2)$ models.
We investigate non-Hermitian generalization of 
the Luttinger Hamiltonian on the square lattice, 
and that of the Kane-Mele model on the honeycomb lattice, respectively.
Using the generalized Kramers theorem for the time-reversal operator
 $\Theta$ with $\Theta^2=+1$ [M. Sato {\it et al.}, arXiv:1106.1806],
we introduce a time-reversal invariant Chern number from which
 topological stability of gapless edge modes is argued.

\end{abstract}

\pacs{73.43.Nq, 73.20.-r, 72.25.-b}

\maketitle

\tableofcontents

\section{Introduction}
\label{sec:introduction}

Recent developments of 
topological insulators have led to much interest in topological phases
of matter.  
Murakami, Nagaosa, and Zhang predicted spin-Hall effect\cite{SM03,SM04_2}
by re-examining Hamiltonians describing hole-doped semiconductors 
with spin-orbit coupling presented 
by Luttinger (the Luttinger Hamiltonian)\cite{Luttinger}.
After the theoretical predictions
and experimental observations\cite{Kato04,Wunderlich}, 
the spin-Hall effects were generalized 
to insulating systems\cite{SM04}.  
Subsequently, the quantum version of the spin Hall effect, the quantum spin Hall effect, was introduced by  Kane and Mele as a model with time-reversal symmetry in graphene \cite{KM05_2} 
and independently by Bernevig and Zhang as two-dimensional semiconductor systems with a uniform strain gradient \cite{BernevigZhang2005}.  
Remarkably, quantum spin-Hall effects were experimentally observed
in 2D CdTe/HgTe/CdTe quantum well by K\"onig {\it et al.}\cite{Konig}, 
 following the theoretical predictions 
by Bernevig, Hughes, and Zhang\cite{BHZ}.

The appearance of the topologically protected gapless edge states within the bulk gap is a manifestation of the topological insulator. 
The number of such gapless edge modes 
is specified by topological invariants.  
In Ref.\cite{QiWuZhang}, gapless edge states carrying spin currents
were found when the spin is conserved. They are characterized by spin
Chern numbers introduced by Sheng {\it et al.}\cite{Sheng}.
The spin Chern number is an extension of the Chern number for quantized Hall conductivity\cite{TKNN,TKNN2} 
to quantized spin Hall conductivity in time-reversal symmetric systems with spin conservation.
Kane and Mele introduced $Z_2$ invariants
that distinguish topologically non-trivial phases 
 from trivial ones in time-reversal symmetric systems\cite{KM05,KM05_2} 
even when the spin is not conserved. 
The above-mentioned gapless edge states in quantum spin Hall insulator, dubbed as the helical edge modes \cite{Wuetal2006}, are known as
a consequence of the Kramers theorem of 
the time-reversal symmetry $\Theta$ with $\Theta^2=-1$.  
In contrast, time-reversal symmetric systems with $\Theta^2=+1$
have not been investigated so far in the context of topological insulators, 
since they are considered to be irrelevant to topologically protected 
gapless edge states.

Recently, the present authors have shown that the (generalized) Kramers
 theorem follows even for $\Theta^2=+1$ in a class of non-Hermitian
 systems (provided the metric operator is anticommutative with the
 time-reversal operator)\cite{HSEK}. Therefore, a natural question
 arises: Do topologically protected edge modes also appear in such
 non-Hermitian models? 
 The main purpose of the present work is to give an answer to the question.  
We demonstrate numerical calculations and provide topological arguments
 for the stability of edge modes in non-Hermitian Hamiltonians. 
In particular, we investigate lattice versions of the $SU(1,1)$ and
$SO(3,2)$ Hamiltonians studied in Ref. \cite{HSEK}. 
(See also the related works, \cite{KH10_1,KH10_2}  in higher dimensional quantum Hall effect, and  \cite{BG11,BG11_2,BG11_3} in {\it PT} symmetric quantum mechanics.) 
As a lattice realization of $SU(1,1)$ model, 
we consider the tight-binding model on the honeycomb lattice with
imaginary on-site potentials.
For $SO(3,2)$ model, we investigate non-Hermitian generalization
of the Luttinger Hamiltonian\cite{SM04,QiWuZhang} on the square lattice. 
We also argue  a non-Hermitian generalization 
of the Kane-Mele model\cite{KM05,KM05_2}, 
where the hopping integrals
are asymmetric due to non-Hermiticity.

Non-Hermitian systems play important roles in physics. 
For instance, a non-Hermitian system with disorder, 
known as the Hatano-Nelson model, 
has been studied in the context of localization-delocalization transition
in one and two dimensions\cite{NH06,NH07,NH08}. 
The model simulates the 
depinning of flux lines
in type-II superconductors subject to a transverse 
magnetic field\cite{NH06}.  
We also briefly discuss possible relevance 
to the model. 

The paper is organized as follows. 
In Sec.\ref{sec:su11so32ham}, we introduce the $SU(1,1)$ and $SO(3,2)$
Hamiltonians and their basic structures. In Sec.\ref{sec:su11_model}, a
lattice version of the $SU(1,1)$ model on graphene is
explored, and the stability of edge states is discussed on the basis of
topological arguments. We further investigate the $SO(3,2)$ model on the square
lattice in Sec.\ref{sec:so32model}. 
A topological number for the non-Hermitian system 
is introduced in order to
account for the stability of edge modes under small non-Hermitian
perturbation. In Sec.\ref{sec:so32}, a non-Hermitian version of the
Kane-Mele model is presented and the topological property of the model
is also discussed. 
Section \ref{sec:summary} is devoted to summary and discussions.  

\section{$SU(1,1)$ and $SO(3,2)$ Hamiltonians}\label{sec:su11so32ham}

In this paper, we mainly consider $SU(1,1)$ and $SO(3,2)$ models,
which are non-Hermitian generalizations of $SU(2)$ and $SO(5)$ models, 
respectively. Here, we briefly discuss their structures with emphasis on the relations to split-quaternions along Ref.\cite{HSEK}.

It is well known that the quaternion algebra is realized as $2\times 2$ unit matrix, $1_2$, and ($i$ times) Pauli matrices, $i\sigma_a$  ($a=1,2,3$).  
The $SU(2)$ Hamiltonian is constructed by the Pauli matrices $\sigma_a$,
\be
H=d_1\sigma_1+d_2\sigma_2+d_3\sigma_3,
\label{su2_d123}
\ee
where $d_a$ ($a=1,2,3$) are real.
The eigenenergies are,
\be
E_{\pm}=\pm\sqrt{d_1^2+d_2^2+d_3^2}.
\ee
Similarly, the split quaternion algebra is realized as the $2\times 2$ unit matrix, $1_2$, and  ($i$ times) $SU(1,1)$ Pauli matrices, $(-\sigma_1,-\sigma_2,i\sigma_3)$, and the $SU(1,1)$ Hamiltonian is constructed by such $SU(1,1)$ Pauli matrices: 
\be
H=d_1\sigma_1+d_2\sigma_2+i d_3\sigma_3.
\label{su11_d123}
\ee
The eigenenergies are,
\be
E_{\pm}=\pm\sqrt{d_1^2+d_2^2-d_3^2}.
\ee

The $SO(5)$ gamma matrices, $\Gamma_a$ ($a=1,2,\ldots,5$),  are  defined so as to satisfy 
the Clifford algebra $\{\Gamma_a,\Gamma_b\}=2\delta_{a b}$. They have the quaternionic structure since their off diagonal components are given by the ``quaternions''
\begin{equation}   
\Gamma_1=\begin{pmatrix}
0 & i\sigma_1\\
-i\sigma_1 & 0 
\end{pmatrix},~~
\Gamma_2=\begin{pmatrix}
0 & i\sigma_2\\
-i\sigma_2 & 0 
\end{pmatrix},~~
\Gamma_3=\begin{pmatrix}
0 & i\sigma_3\\
-i\sigma_3 & 0 
\end{pmatrix},~~
\Gamma_4=\begin{pmatrix}
0 & 1_2\\
1_2 & 0 
\end{pmatrix},~~
\Gamma_5=\begin{pmatrix}
1_2 & 0\\
0 & -1_2 
\end{pmatrix}. 
\label{so5gammama}
\end{equation}
With the $SO(5)$ gamma matrices, 
the $SO(5)$ Hamiltonian is given by 
\be
H=d_1\Gamma_1+d_2\Gamma_2+d_3\Gamma_3+d_4\Gamma_4+d_5\Gamma_5,
\ee
where $d_a$ ($a=1,2,\ldots,5$) are real.
The eigenenergies are,
\be
E_{\pm}=\pm\sqrt{d_1^2+d_2^2+d_3^2+d_4^2+d_5^2},
\ee
each of which is doubly degenerate. Such double degeneracy is understood as a consequence of the Kramers theorem since the $SO(5)$ Hamiltonian is invariant under time-reversal operation $\Theta_-^2=-1$ with  
\begin{equation}
\Theta_-=
\begin{pmatrix}
i\sigma_2 & 0 \\
0 & i\sigma_2
\end{pmatrix}\cdot K, 
\end{equation}
where $K$ denotes the complex conjugation operator. 
Similarly, the $SO(3,2)$ gamma matrices are introduced as the matrices whose off-diagonal blocks are given by the split-quaternions. Hence, the $SO(3,2)$ gamma matrices are given by 
\begin{equation}   
\begin{pmatrix}
0 & -\sigma_1\\
\sigma_1 & 0 
\end{pmatrix},~~
\begin{pmatrix}
0 & -\sigma_2\\
\sigma_2 & 0 
\end{pmatrix},~~
\begin{pmatrix}
0 & i\sigma_3\\
-i\sigma_3 & 0 
\end{pmatrix},~~
\begin{pmatrix}
0 & 1_2\\
1_2 & 0 
\end{pmatrix},~~
\begin{pmatrix}
1_2 & 0\\
0 & -1_2 
\end{pmatrix},  
\label{so32gammam}
\end{equation}
which are 
$i\Gamma_1,i\Gamma_2,\Gamma_3,\Gamma_4,\Gamma_5$ with the $SO(5)$ gamma matrices (\ref{so5gammama}). From the $SO(3,2)$ gamma matrices, we construct the $SO(3,2)$
Hamiltonian as  
\be
H=i d_1\Gamma_1+i d_2\Gamma_2+d_3\Gamma_3+d_4\Gamma_4+d_5\Gamma_5.
\ee
The eigenenergies are,
\be
E_{\pm}=\pm\sqrt{d_3^2+d_4^2+d_5^2-d_1^2-d_2^2},
\ee
each of which is doubly degenerate. 
The $SO(3,2)$ Hamiltonian is invariant under time-reversal operation $\Theta_+^2=+1$ with  
\begin{equation}
\Theta_+=
\begin{pmatrix}
\sigma_1 & 0 \\
0 & \sigma_1
\end{pmatrix}\cdot K. 
\end{equation}
As discussed in Ref.\cite{HSEK}, such double degeneracy is a consequence of the generalized Kramers theorem for $\Theta^2_+=+1$.  The metric operator that satisfies $\eta H \eta^{-1}=H^{\dagger}$ is given by 
\begin{equation}
\eta=
\begin{pmatrix}
\sigma_3 & 0 \\
0 & \sigma_3
\end{pmatrix}. 
\end{equation}
Apparently,  $\eta$ is anticommutative with $\Theta_+$: $\{\eta,\Theta_+\}=0$. 

\section{$SU(1,1)$ model and winding numbers}
\label{sec:su11_model}

In this section, we investigate topological stability of the edge states 
of the $SU(1,1)$ model. As an example, 
the tight-binding model on the honeycomb lattice with imaginary on-site
potentials 
is considered. 
It will be found that
edge states with Re$E=0$ are robust under weak non-Hermiticity. 
We will also argue that their topological stability is guaranteed by
topological reasoning.

\subsection{Graphene with imaginary sublattice potential}
\begin{figure}
 \begin{center}
  \includegraphics[width=6.0cm,clip]{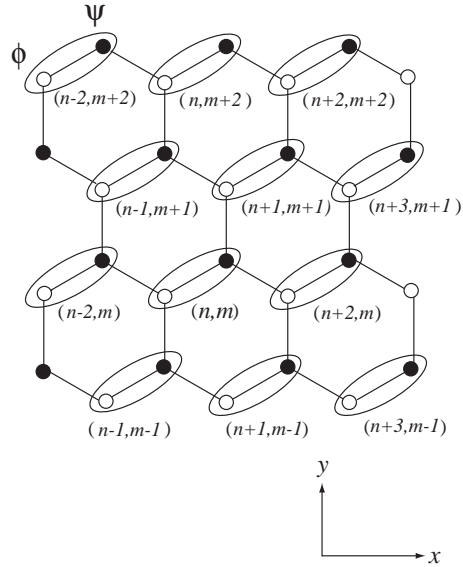}
\caption{\label{fig:honeycomb_lattice_QSHE}
 The honeycomb lattice. }
\end{center}
\end{figure} 
\begin{figure}
 \begin{center}
  \includegraphics[width=8.0cm,clip]{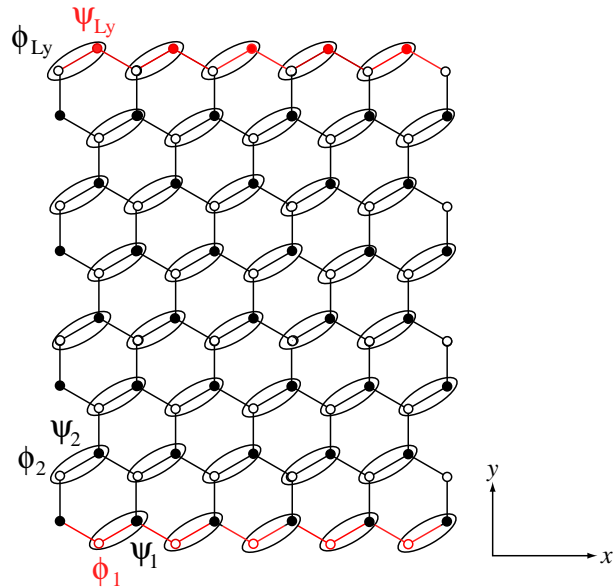}
\caption{\label{fig:honeycomb_lattice_zigzag}
The honeycomb lattice with zigzag edges at $m=1$ and $m=8$ ($L_y=8$)
  along the $x$-direction.
}
\end{center}
\end{figure} 

As a model with the $SU(1,1)$ Hamiltonian,
we consider the following tight-binding model on the honeycomb lattice:
\be
H=t\sum_{\langle i,j\rangle} (c_i^{\dag} c_j+{\rm H.c.}) -i \sum_i \lambda_i c_i^{\dag}c_i,
\quad (\lambda_i>0).
\label{su11_pre}
\ee
The first term is the nearest-neighbor hopping.
The second term is imaginary on-site sublattice potentials
which make the system non-Hermitian.
Here, $\lambda_i=\lambda_{a}$ and $\lambda_{b}$
for closed and open circles in Fig.\ref{fig:honeycomb_lattice_QSHE},
respectively.
The Hamiltonian (\ref{su11_pre}) is rewritten as
\be
H=t\sum_{\langle i,j\rangle} 
(c_i^{\dag} c_j+{\rm H.c.}) +i \lambda_V\sum_i \xi_i c_i^{\dag}c_i
-i\frac{\lambda_a+\lambda_b}{2} \sum_i c_i^{\dag}c_i,
\ee
where $\lambda_V=\frac{\lambda_b-\lambda_a}{2}$ and 
$\xi_i=+1$ ($-1$) for closed (open) circles 
in Fig.\ref{fig:honeycomb_lattice_QSHE}. 
Since the $-i\frac{\lambda_a+\lambda_b}{2} \sum_i c_i^{\dag}c_i$ term
only shifts the origin of the energy, we neglect it in the following:
\be
H=t\sum_{\langle i,j\rangle} (c_i^{\dag} c_j+{\rm H.c.}) 
+i \lambda_V\sum_i \xi_i c_i^{\dag}c_i.
\label{su11_v}
\ee

In the absence of edges, (\ref{su11_v}) 
reduces to the following $SU(1,1)$ Hamiltonian in the momentum space:
\be
H({\bm k})=\left(
\begin{array}{cc} 
    i\lambda_V & 2t \cos\left(\frac{k_x a}{2}\right)+t e^{i\frac{\sqrt{3}k_y a}{2}} \\ 
   2t \cos\left(\frac{k_x a}{2}\right)+t e^{-i \frac{\sqrt{3} k_y a}{2}}
 & -i\lambda_V
\end{array} \right),
\label{graphene}
\ee
which is obtained by the Fourier transformations,
\be
c_i&=&\psi_{(n,m)}=\sum_{{\bm k} } e^{i \frac{a}{2} k_x n+i \frac{\sqrt{3}a}{2}k_y m} 
\psi_{\bm k},\nonumber \\
c_{i+{\bm d}}&=&\phi_{(n+1,m+1)}=\sum_{{\bm k}} e^{i \frac{a}{2} k_x n+i
\frac{\sqrt{3}a}{2}k_y (m+1)} \phi_{\bm k}.
\ee
Here, $\psi_{\bm k}$ and $\phi_{\bm k}$ are the annihilation operators 
in the momentum space corresponding to 
the closed and the open circles 
in Fig.\ref{fig:honeycomb_lattice_QSHE}, respectively, 
 $k_x$ ($k_y$) the momentum in the $x$ ($y$) direction,
$a$  the lattice constant of the honeycomb lattice, 
and ${\bm d}=a {\bm {\hat y}}/\sqrt{3}$.
The bulk spectra of the system are obtained by diagonalizing the
Hamiltonian (\ref{graphene}),
\be
E_{\pm}({\bm k})=\pm\sqrt{\left|t+2 t\cos\left(\frac{k_x a}{2}\right)
e^{i \frac{\sqrt{3} k_y a}{2}}\right|^2-\lambda_V^2}.
\label{su11_energy}
\ee

\begin{figure}
 \begin{center}
\includegraphics[width=6.0cm,clip]
{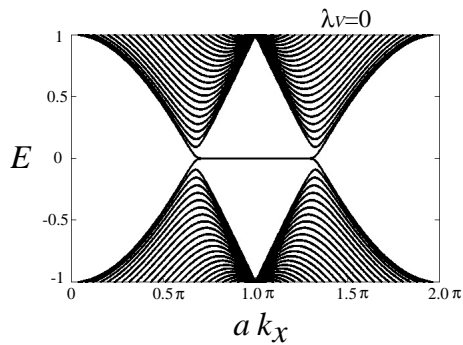}
\caption{The energy bands of the $SU(1,1)$ model (\ref{su11_v}) with
  zigzag edges along the $x$-direction 
for $t=1.0$ and $\lambda_V=0$ (Hermitian case).
Here $a$ is the lattice constant and $k_x$ the momentum in the $x$-direction.
A zero energy edge state with flat band appears for $2\pi/3<ak_x<4\pi/3$.}
\label{fig:su2}
\end{center}
\end{figure}

\begin{figure}
 \begin{center}
\includegraphics[width=8.5cm,clip]
{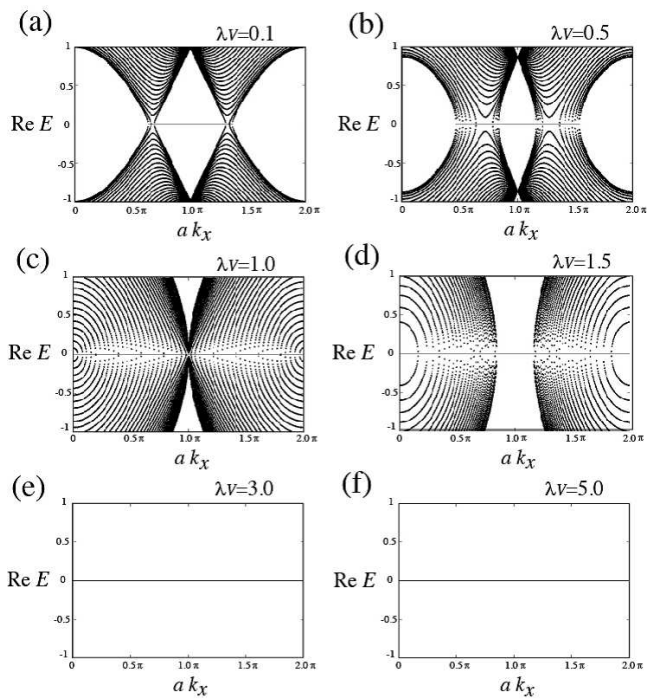}
\caption{The real part of the energy bands of the $SU(1,1)$ model
  (\ref{su11_v}) with zigzag edges along the $x$-direction.
We show the results for  $t=1.0$ and various values of non-Hermitian
  parameter $\lambda_V$. 
Here $a$ is the lattice constant and $k_x$ the momentum in the $x$-direction.
For weak non-Hermiticity [(a) and (b)], the edge
  state with ${\rm Re}E=0$ survives around $ak_x\sim \pi$.
For large $\lambda_V$ [(c)-(f)], no edge state with ${\rm Re}E=0$ appears.}
\label{fig:su11_potential}
\end{center}
\end{figure}

\begin{figure}
 \begin{center}
\includegraphics[width=8.5cm,clip]
{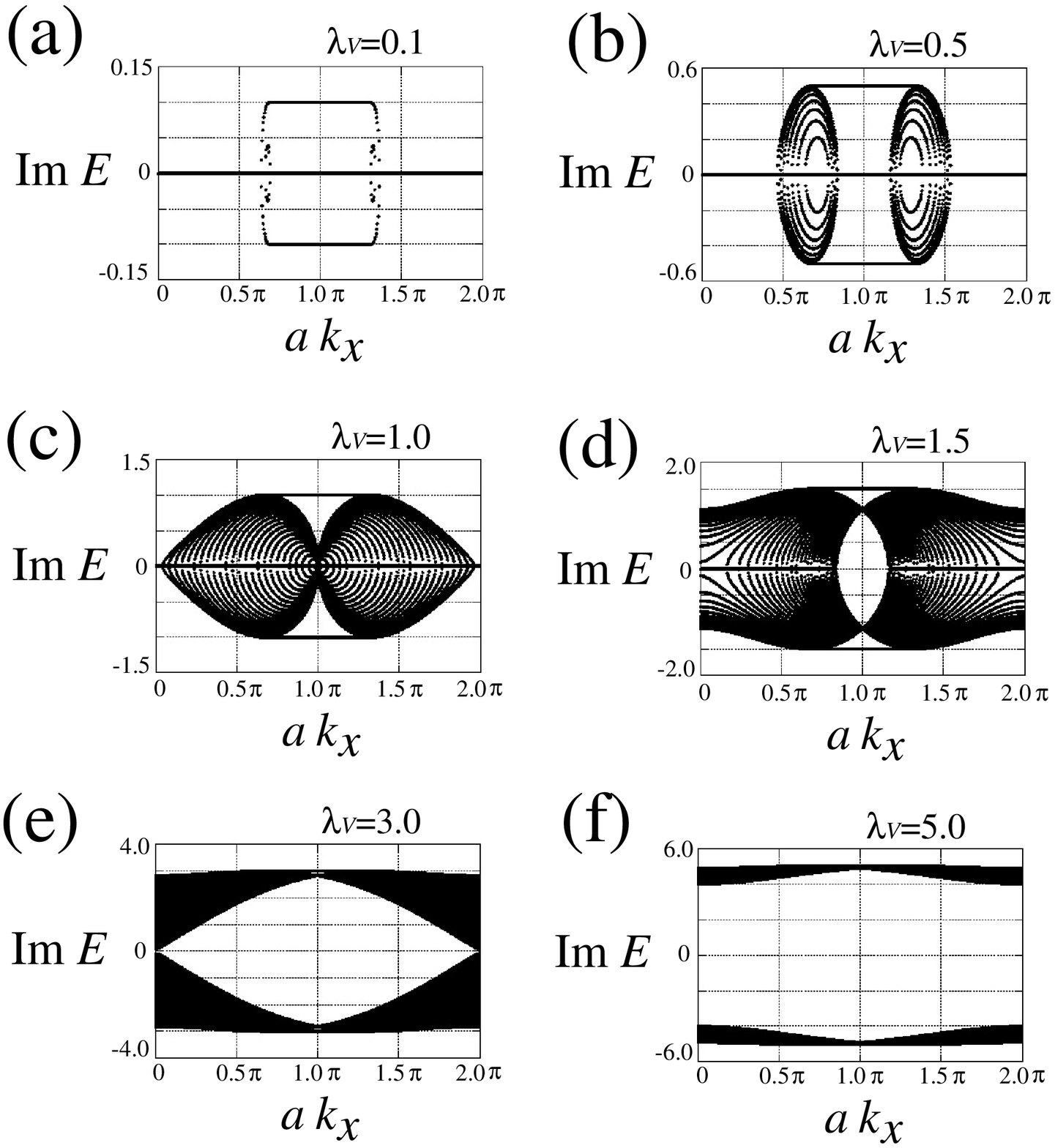}
\caption{The imaginary part of the energy bands of the $SU(1,1)$ model
  (\ref{su11_v}) with zigzag edges along the $x$-direction. We plot the
  results for  $t=1.0$ and various values of the non-Hermitian parameter
  $\lambda_V$.
Here $a$ is the lattice constant and $k_x$ the momentum in the $x$-direction.
For weak non-Hermiticity [(a) and (b)], there is no imaginary part in
  the bulk except near the gap closing points at $a k_x=2\pi/3$ and
  $ak_x=4\pi/3$. 
The edge states with ${\rm Re}E=0$
  [Fig.\ref{fig:su11_potential} (a) and (b)] support the imaginary part
  in their energy. 
For large $\lambda_{V}$ [(c)-(f)], the bulk states support the
  imaginary part. }
\label{fig:su11_potential2}
\end{center}
\end{figure}

Let us now examine the edge state in this system.
If we make zigzag edges along the $x$ direction 
as shown in Fig.\ref{fig:honeycomb_lattice_zigzag}, 
a zero energy edge band appears in addition to the bulk bands
(\ref{su11_energy}): 
For the Hermitian case $\lambda_V=0$, it has been known that the edge state
with $E=0$ appears for $2\pi/3<a k_x < 4\pi/3$ as shown in
Fig.\ref{fig:su2} \cite{Kusakabe}. 
We find here that the zero energy edge state persists even in the
presence of weak
non-Hermiticity $\lambda_V$.
(See Figs.\ref{fig:su11_potential} and \ref{fig:su11_potential2}.)
We illustrate the real part and the imaginary part of the energy bands
 as functions of $a k_x$
in Fig.\ref{fig:su11_potential} and  Fig.\ref{fig:su11_potential2},
respectively.  
It is clearly seen that the edge state with Re $E=0$ persists 
in the region $2\pi/3<a k_x<4\pi/3$ for small $\lambda_V$.


\subsection{Winding number and generalized Index theorem for
  non-hermitian system}
\label{sec:winding_number}

\begin{figure}
 \begin{center}
\includegraphics[width=8.5cm,clip]{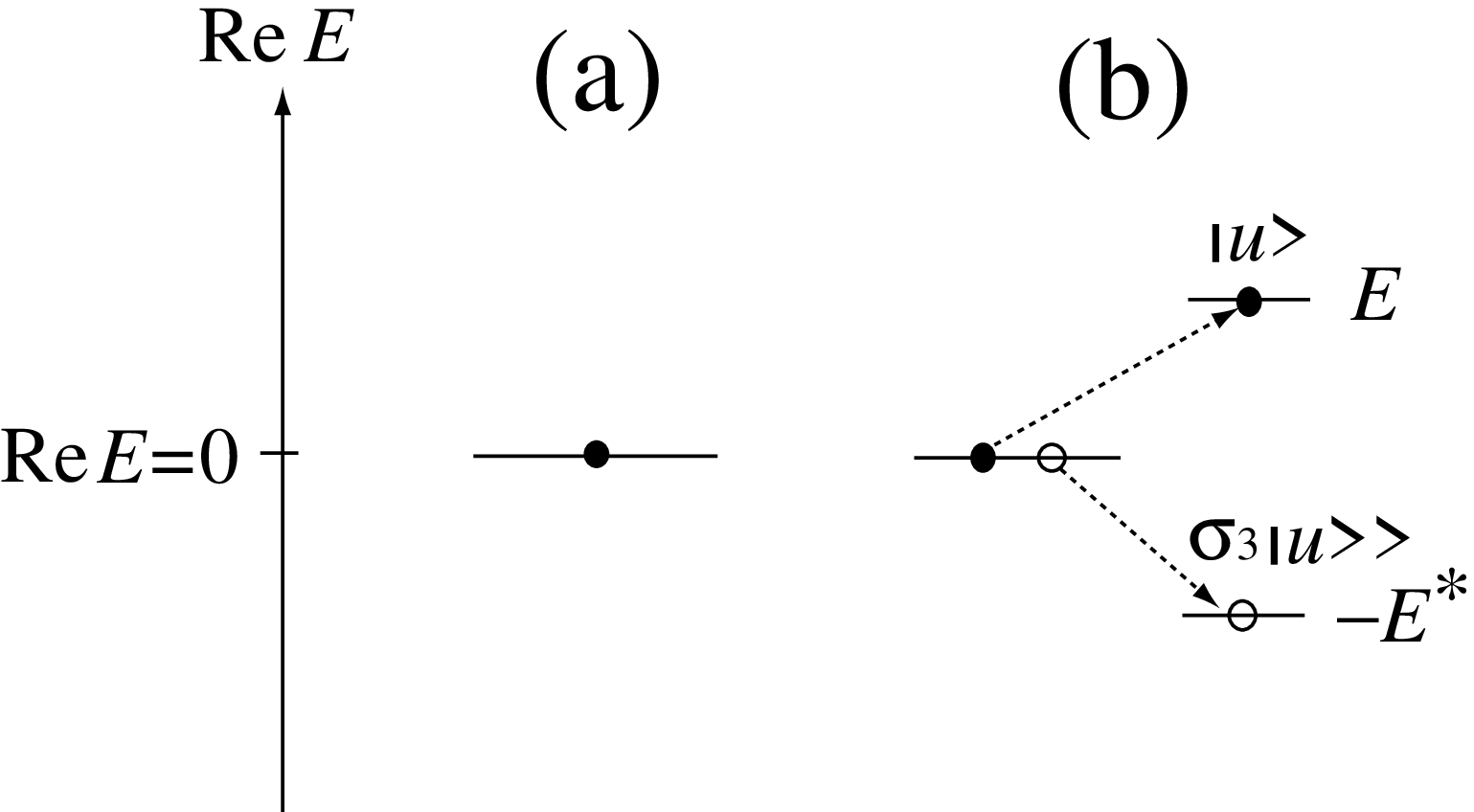}
\caption{(a) Topologically protected  state with Re$E=0$.
(b) Topologically trivial states with Re$E=0$.
Here, the closed (open) circles at Re$E=0$ represent states
satisfying $\sigma_3 |u \rangle\!\rangle=+\ket{u}$ (\ref{su11_rel1})
($\sigma_3 |{u}\rangle\!\rangle=-\ket{u}$ (\ref{su11_rel2})). In the
  latter case [(b)],
  the state can be non-zero mode since
  $n_+^0-n_-^0=0$. }
\label{fig:chiral_zero}
\end{center}
\end{figure} 

From the bulk-edge correspondence, the zero energy edge state
suggests the existence of a topological number responsible for the edge mode.
In this section, we will see that this is indeed the case and the edge
state is characterized by the one-dimensional winding number. 

\subsubsection{Basic property of edge states}
\label{sec:index}
 
In order to make our arguments concrete, we consider the semi-infinite
$SU(1,1)$ model on $m>0$.
We perform the inverse Fourier transformation of $H({\bm k})$ in
(\ref{graphene}) with respect to $k_y$ and denote the resultant Hamiltonian as
$H(k_x)_{m, m'}$.
The energy spectrum of the system is given by 
\begin{eqnarray}
\sum_{m'}H(k_x)_{m,m'}|u(k_x,m')\rangle=E(k_x)|u(k_x, m)\rangle,
\label{eq:edgeH}
\end{eqnarray}
with 
\begin{eqnarray}
|u(k_x, m) \rangle=0,
\label{eq:boundarycond}
\end{eqnarray}
for $m\le 0$.
In the following, we consider $k_x$ as a parameter of the system and
treat the model as a one-dimensional system along the $y$-direction. 
We also restrict our argument on $k_x$ where the bulk energy gap is open
in the real part of the energy. 
The zero energy edge state satisfies (\ref{eq:edgeH}) and
(\ref{eq:boundarycond}) with ${\rm Re}E(k_x)=0$.

As was shown in Ref.\cite{HSEK}, the basic symmetry of the $SU(1,1)$
Hamiltonian is pseudo-anti-Hermiticity given by
\be
\sigma_3 H({\bm k})^{\dag}\sigma_3=-H({\bm k}),
\label{sym_su11_k}
\ee
which also implies the pseudo-anti-Hermiticity of the Fourier transformed
one, 
\be
\sigma_3 [H(k_x)^{\dag}]_{m,m'}\sigma_3=-H(k_x)_{m,m'}.
\label{sym_su11}
\ee
The pseudo-anti-Hermiticity is a key ingredient of our topological argument.

Let us denote the right eigenvectors and the left eigenvectors of
$H(k_x)_{m,m'}$  as $\ket{u(k_x, m)}$ and $|u(k_x, m)\rangle\!\rangle$,
respectively: 
\be
\sum_{m'}H(k_x)_{m,m'}\ket{u(k_x, m')}=E(k_x)\ket{u(k_x, m)},
\nonumber\\
\sum_{m'}[H(k_x)^{\dag}]_{m,m'}|u(k_x, m')\rangle\!\rangle 
=E(k_x)^*|u(k_x, m)\rangle\!\rangle, 
\ee
where $\ket{u(k_x, m)}$ and $|u(k_x, m)\rangle\!\rangle$ are normalized as 
\be
\sum_m\langle\!\langle u(k_x, m)|u(k_x, m) \rangle
=\sum_m\langle u(k_x, m)| u(k_x, m)\rangle\!\rangle=1.
\label{normalize_su11_u}
\ee
From the pseudo-anti-Hermiticity (\ref{sym_su11}), we find that 
\begin{eqnarray}
\sum_{m'}H(k_x)_{m,m'}\sigma_3|u(k_x, m')\rangle\!\rangle=
- E(k_x)^*\sigma_3 |u(k_x, m)\rangle\!\rangle.
\end{eqnarray}
Therefore, the eigenstate
$\ket{u(k_x, m)}$ with eigenenergy $E(k_x)$ always comes in pairs with
the eigenstate $\sigma_3 |u(k_x, m) \rangle\!\rangle $ with eigenenergy
$-E(k_x)^{*}$. 
When Re$E(k_x)\ne 0$, $\ket{u(k_x,m)}$ and $\sigma_3 |u(k_x, m)
\rangle\!\rangle$  are independent of each other, since they have
different energies.
On the other hand, for edge states with Re$E(k_x)=0$, they are not
always independent.
Actually, by choosing a proper basis, they can be related to each other as
\be
\sigma_3  |u(k_x, m) \rangle\!\rangle 
=+ \ket{u(k_x, m)}, 
\label{su11_rel1} 
\ee
or
\be
\sigma_3  |u(k_x, m) \rangle\!\rangle = - \ket{u(k_x, m)},
\label{su11_rel2}
\ee
where the overall phase factors of the right-hand sides 
of (\ref{su11_rel1}) and (\ref{su11_rel2}) are restricted to
$\pm 1$ due to the normalization condition (\ref{normalize_su11_u}).
We denote the number of the edge states with Re$E(k_x)=0$ satisfying
(\ref{su11_rel1}) and (\ref{su11_rel2}) as $n_{+}^{0}$ and $n_{-}^{0}$,
respectively. 

Here we can show an important property of the edge states: The difference
between $n_{+}^{0}$ and $n_{-}^{0}$ does not change its value against
perturbation preserving the pseudo-anti-Hermiticity. 
The reason why $n_{+}^{0}-n_{-}^{0}$ is conserved is as follows.
For Re$E(k_x)\ne 0$, $\ket{u(k_x, m)}$ and $\sigma_3 |u(k_x, m)
\rangle\!\rangle$ are independent,
so we can construct another basis from them as
\be
\ket{u_{\pm}(k_x, m)}
=\frac{1}{\sqrt{2}}( |u(k_x, m) \rangle\!\rangle 
\pm \sigma_3\ket{u(k_x,m)}),
\nonumber\\
|{u}_{\pm}(k_x, m)\rangle\!\rangle
=\frac{1}{\sqrt{2}}(\ket{u(k_x, m)}\pm \sigma_3  |u(k_x, m) \rangle\!\rangle),
\ee
with
\be
  \langle\!\langle {u}_{\pm}(k_x, m) |u_{\pm}(k_x, m)\rangle=1,
\quad 
\langle\!\langle {u}_{\pm}(k_x, m) |u_{\mp}(k_x, m)\rangle=0.
\ee
Since these states $\ket{u_{+}(k_x, m)}$ and $\ket{u_{-}(k_x, m)}$
satisfy (\ref{su11_rel1}) and (\ref{su11_rel2}), respectively
\be
\sigma_3  | {u}_{\pm}(k_x, m) \rangle\!\rangle=\pm  \ket{u_{\pm}(k_x,
m)}, 
\label{nonzero_rel}
\ee
we have the same number of states with signs
$+$ and $-$ in the right-hand side of (\ref{nonzero_rel}) for ${\rm
Re}E(k_x)\neq 0$.  
This means that $n^0_+-n_-^0$ can not change adiabatically:
By small perturbation, some of the edge states may acquire a non-zero
real part of the energy.
If this happens, however, they must be paired with opposite sign in the right
hand side of (\ref{nonzero_rel}).
As a result, the difference between $n^0_+$ and $n_-^0$ does not change
at all.
In Fig.\ref{fig:chiral_zero}, we illustrate two different edge states
with Re$E=0$, one is topologically protected from the above
argument and the other topologically trivial.

\subsubsection{Winding number}
\label{sec:winding}

Now we introduce a bulk topological number relevant to the present edge mode.
To do this, consider the eigenequation for $H({\bm k})$
in the momentum space:
\be
H({\bm k})\ket{u_n({\bm k})}=E_n({\bm k})\ket{u_n({\bm k})},\quad
H({\bm k})^{\dag} |u_n({\bm k})\rangle\!\rangle=E_n({\bm k})^* 
|{u}_n({\bm k})\rangle\!\rangle,
\ee
where $n$ denotes the index labeling different bands.
We assume that the system is half-filling and that
we have a gap in the real part of energy around Re$E({\bm k})=0$ for a
fixed value of $k_x$.

From the pseudo-anti-Hermiticity (\ref{sym_su11_k}), one can say that if
$|u_n({\bm k})\rangle $ is a right eigenstate of $H({\bm k})$ with ${\rm Re}E_n({\bm k})>0$,
$\sigma_3|u_n({\bm k})\rangle$ is a left eigenstate of $H({\bm k})$ with 
${\rm Re}E_n({\bm k})<0$.
In the following, we use a positive (negative) $n$ for the state with 
${\rm Re}E_n({\bm k})>0$ (${\rm Re}E_n({\bm k})<0$), and set the
relation
\be
|{u}_{-n}({\bm k}) \rangle\!\rangle=
\sigma_3 \ket{u_{n}({\bm k})}, 
\quad 
E_{-n}({\bm k})^*=-E_n({\bm k}).
\label{su11_gamma_rel}
\ee

The topological number is constructed from the non-Hermitian
generalization 
of the projection operators $\widetilde{\cal{P}}_1({\bm k})$
and $\widetilde{\cal{P}}_2({\bm k})$ for occupied bands \cite{Ryu_Furusaki08,Avron83}:
\be
&&\widetilde{\cal{P}}_1({\bm k})
=\sum_{n<0} |{u}_n ({\bm k})\rangle\!\rangle\bra{u_n
({\bm k})},
\quad
\widetilde{\cal{P}}_2({\bm k})
=\sum_{n<0} \ket{u_n ({\bm k})}\langle\!\langle {u}_n
({\bm k})|,
\nonumber\\
&&\left(\widetilde{\cal{P}}_1({\bm
k})\right)^2=\widetilde{\cal{P}}_1(\bm k),\quad
\left(\widetilde{\cal{P}}_2({\bm k})\right)^2=\widetilde{\cal{P}}_2({\bm
k}).
\ee
From $\widetilde{{\cal P}}_1({\bm k})$ and $\widetilde{{\cal
P}}_2({\bm k})$, 
we define the following $\cal{Q}$ matrix:
\be
\widetilde{\cal{Q}}({\bm k})
&=&{\bm 1}-\left[\widetilde{\cal{P}}_1({\bm
k})+\widetilde{\cal{P}}_2({\bm k})\right]
 \nonumber\\
&=& \frac{1}{2} 
 \left(\sum_{n>0} | {u}_n({\bm k})\rangle\!\rangle\bra{u_n({\bm k})}
-\sum_{n<0} |{u}_n({\bm k})\rangle\!\rangle \bra{u_n({\bm k})} \nonumber  \right. \\
&{}&\left. +  \sum_{n>0} \ket{u_n({\bm k})}
\langle\!\langle {u}_n({\bm k})|-\sum_{n <0} 
\ket{u_n({\bm k})}\langle\!\langle {u}_n({\bm k})| \right),
\label{Q_expression}
\ee
where the completeness relation 
$\sum_{n} |{u}_n({\bm k})\rangle\!\rangle \bra{u_n({\bm k})}=
\sum_{n} \ket{u_n({\bm k})} \langle\!\langle {u}_n({\bm k})|
={\bm 1}$ was used.
One can immediately show that the matrix $\widetilde{\cal{Q}}({\bm k})$ is
Hermitian:
\be
\widetilde{\cal{Q}}({\bm k})^{\dag}=\widetilde{\cal{Q}}({\bm k}).
\label{su11_Qdag}
\ee
From (\ref{su11_gamma_rel}) and (\ref{Q_expression}), 
we find that
$\widetilde{\cal{Q}}({\bm k})$ and $\sigma_3$ anticommute:
\be
\{\widetilde{\cal{Q}}({\bm k}),\sigma_3 \}=0.
\label{su11_anti}
\ee
Since $\sigma_3$ is diagonal, (\ref{su11_anti}) implies that  
$\widetilde{\cal{Q}}({\bm k})$ is off-diagonal and can be expressed by a
complex function $q({\bm k})$ as 
\be
\widetilde{\cal{Q}}({\bm k})
=\left(
\begin{array}{cc} 
  0 & q({\bm k}) \\ 
  q({\bm k})^* & 0
\end{array} \right).
\ee

The topological number relevant to our model is the one-dimensional
winding number $w_{\rm 1d}$ defined as 
\be
w_{\rm 1d}(k_x)
&=& \frac{1}{2\pi i} \int_{0}^{2\pi} q({\bm
k})^{-1}\frac{\partial}{\partial k_y}
q({\bm k}) dk_y .
\label{winding_1d}
\ee
By using polar coordinates $q({\bm k})=|q({\bm k})|e^{i\alpha({\bm
k})}$, we obtain,
\be
w_{\rm 1d}(k_x)&=&\frac{1}{2\pi}\int_{0}^{2\pi} 
\frac{\partial\alpha({\bm k})}{\partial k_y} dk_y
  +\frac{1}{2\pi i}\int_{0}^{2\pi} \frac{\partial}{\partial k_y} 
[\ln |q({\bm k})|] dk_y,\nonumber\\
&=& N, \quad (N: {\rm integer}),
\ee
where we have used the periodicity of $q({\bm k})$ with respect to $k_y$
to obtain a quantized value $N$.
The bulk-edge correspondence implies that if the winding number $w_{\rm
1d}(k_x)$ is nonzero, there is a zero energy state (in the real
part of the energy) on the boundary.
We will confirm this numerically in Sec.\ref{sec:topological}.

\subsubsection{Generalized Index theorem}

In Sec.\ref{sec:winding}, we argue that the non-zero $w_{\rm 1d}(k_x)$
implies the existence of edge states with ${\rm Re}E=0$.
At the same time, in Sec.\ref{sec:index}, we find that the non-zero
$n_+^0-n_-^0$ ensures the robustness of the existence of edge states
with ${\rm Re}E=0$. 
Therefore, it is natural to identify these two quantities $w_{\rm 1d}(k_x)$ 
and $n_+^0-n_-^0$. 
Since there is a sign ambiguity for the identification, two
possible relations are suggested in the form of the index theorem,
\begin{eqnarray}
n^0_--n^0_+=w_{\rm 1d}(k_x), 
\label{eq:index1}
\end{eqnarray}
or
\begin{eqnarray}
n^0_+-n^0_-=w_{\rm 1d}(k_x). 
\label{eq:index2}
\end{eqnarray}
Here note that there are two possible choices of the edge of the system,
{\it i.e.}, the surface of the semi-infinite system on $y>0$ or that on
$y<0$.
The two possible choices of the surface correspond to two possible
equalities (\ref{eq:index1}) and (\ref{eq:index2}).

In this paper, we will not prove the generalized index theorem
(\ref{eq:index1}) and (\ref{eq:index2}).
In Ref.\cite{STYY11}, one of the present authors proved a similar
generalized index theorem for zero energy edge states in systems with
chiral symmetry.  
We expect that the generalized index theorem in this case can be proved
in a similar manner.

\subsection{Application to $SU(1,1)$ Hamiltonian} 
\label{sec:topological}

Here we present explicit calculations of the winding number $w_{\rm 1d}(k_x)$
for our $SU(1,1)$ Hamiltonian (\ref{graphene}).
Since we treat $k_x$ as a fixed parameter of the system, 
it is convenient to write (\ref{graphene}) as
\begin{eqnarray}
H({\bm k})=\left(
\begin{array}{cc} 
  i\gamma & v+v'e^{i k}\\ 
   v+v'e^{-i k} & -i\gamma
\end{array} \right), 
\label{su11}
\end{eqnarray}
with the identification
\be
\gamma=\lambda_V, \quad v=2t\cos\left(\frac{k_x a}{2}\right), \quad
v'=t, \quad k=\frac{\sqrt{3}k_y a}{2}.
\ee

Let us write (\ref{Q_expression}) as
\be
\widetilde{{\cal Q}}({\bm k})=\widetilde{{\cal Q'}}({\bm k})
+\widetilde{{\cal Q'}}({\bm k})^{\dag},
\label{Q_Qdag}
\ee
where
\be
\widetilde{{\cal Q'}}({\bm k})&=&\frac{1}{2} 
 \left(\sum_{n>0}|{u}_n({\bm k}) \rangle\!\rangle \bra{u_n({\bm
 k})}-\sum_{n <0}
|{u}_n({\bm k}) \rangle\!\rangle\bra{u_n({\bm k})}\right).
\ee
Then, straightforward diagonalization of (\ref{su11}) gives
\be
\widetilde{{\cal Q'}}({\bm k})
&=&\frac{1}{2 \sqrt{|v+v'e^{ik}|^2-\gamma^2}}
\left(
\begin{array}{cc} 
  -i\gamma & v+v'e^{i k}\\ 
   v+v'e^{-i k} & i\gamma
\end{array} \right)\nonumber\\
&=&\frac{1}{2\sqrt{|v+v'e^{ik}|^2-\gamma^2}}H({\bm k})^{\dag}.
\label{Q_explicit}
\ee
From (\ref{Q_Qdag}) and (\ref{Q_explicit}), we obtain
\be
\widetilde{{\cal Q}}({\bm k})
&=&\frac{1}{2 \sqrt{|v+v'e^{ik}|^2-\gamma^2}}[H({\bm k})^{\dag}+H({\bm
k})]\nonumber\\ 
&=&\frac{1}{\sqrt{|v+v'e^{ik}|^2-\gamma^2}}
\left(
\begin{array}{cc} 
      0     & v+v'e^{i k}\\ 
   v+v'e^{-i k} & 0
\end{array} \right).
\ee
Therefore, the winding number $w_{\rm 1d}(k_x)$ is evaluated by
(\ref{winding_1d}) with
$q({\bm k})=\frac{1}{\sqrt{|v+v'e^{ik}|^2-\gamma^2}}(v+v'e^{i k})$. 
Since only the $(v+v'e^{i k})$ part contributes to
the winding number, we have
\be
w_{\rm 1d}(k_x)=\frac{v'}{2\pi}\int_{0}^{2\pi} \frac{e^{ik}}{v+v'e^{ik}}dk.
\ee
Furthermore, changing variable as $z=e^{ik}$ leads to
\be
w_{\rm 1d}(k_x)=\frac{v'}{2\pi i}\oint \frac{1}{v+v'z}dz,
\ee
where the integral is taken over a unit circle $|z|=1$.
Using the residue theorem, we reach the final result, 
\be
w_{\rm 1d}(k_x)&=& 1,\quad (|v'|> |v|),\nonumber\\
        &=& 0,\quad (|v'|< |v|).
\label{windingnumber}
\ee
Therefore, the bulk-edge correspondence predicts the existence of the edge
state with ${\rm Re}E=0$ if $|v'|>|v|$.

In terms of the original parameters of the Hamiltonian (\ref{graphene}),  
the inequality $|v'|>|v|$ means
\be  
2\left|\cos\left(\frac{k_x a}{2}\right)\right|< 1.
\label{criterion}
\ee
Thus 
the edge state with ${\rm Re}E=0$ is predicted 
in a region of  $2\pi/3<ak_x<4\pi/3$, provided $k_x$ also satisfies
$\left(2t |\cos\left(\frac{k_x a}{2}\right)|-t \right)^2>\lambda_V^2$
so that the bulk energy gap is open.
This prediction is clearly confirmed 
in Figs. \ref{fig:su11_potential} and \ref{fig:su11_potential2}.

\section{$SO(3,2)$ model and Time-reversal invariant Chern number
}
\label{sec:so32model}

In this section, we examine $SO(3,2)$ Hamiltonians as a higher symmetric
generalization of the $SU(1,1)$ ones examined in Sec. \ref{sec:su11_model}.
As a concrete example,  
we consider a non-Hermitian generalization
of the Luttinger Hamiltonian\cite{SM04,QiWuZhang} on the square lattice.
We find that gapless edge states exist under weak non-Hermiticity.
To explain the topological origin of the gapless edge states, we introduce 
a time-reversal invariant Chern number inherent to a class of non-Hermitian
Hamiltonians, based on discrete symmetry of the system.
From the bulk-edge correspondence, the time-reversal invariant Chern
number ensures the existence of the gapless edge states in the
non-Hermitian system.

\subsection{General Hamiltonian with time-reversal symmetry 
and pseudo-Hermiticity}
\label{sec:so32_pre}

We first derive a general $4\times 4$ non-Hermitian Hamiltonian
with pseudo-Hermiticity \cite{pseudo,pseudo2},  
which is invariant under time-reversal symmetry $\Theta_+$ 
with $\Theta_{+}^2=+1$ as well.
Such a Hamiltonian
belongs to one of the 43 classes of random matrix
 classification for non-Hermitian systems presented in Ref.\cite{Bernard}.
Gapless edge states obtained in the following are expected to be
robust against disorder in the same class.

A general $4\times 4$ complex Hamiltonian $H({\bm k})$ 
can be represented by a linear combination of
 the identity matrix, 5 gamma matrices $\Gamma_a$, 
and 10 commutators $\Gamma_{ab}=[\Gamma_a,\Gamma_b]/(2i)$ $(a,
b=1,2,\cdots, 5)$:
\be
H({\bm k})=h_0({\bm k})+\sum_{a=1}^{5} h_a({\bm k}) \Gamma_a
+\sum_{a<b=1}^{5} h_{ab}({\bm k}) \Gamma_{ab},
\label{hexp}
\ee
where $h_0({\bm k})$, $h_a({\bm k})$'s and $h_{ab}({\bm k})$'s 
are complex functions of ${\bm k}$. 
We adopt the following representation of the gamma matrices:
\be
\Gamma_1=
\left(
\begin{array}{cc} 
    0 & 1_{2}\\ 
    1_{2} & 0
\end{array} \right),\quad
\Gamma_2=
\left(
\begin{array}{cc} 
    0 & -i 1_{2} \\ 
    i 1_{2} & 0
\end{array} \right),\quad
\Gamma_{\alpha}=
\left(
\begin{array}{cc} 
    \sigma^{\alpha-2} & 0\\ 
    0 &  -\sigma^{\alpha-2}
\end{array} \right),
\label{gammaso32}
\ee
where $\alpha=3,4,5$ and $\sigma^{\mu}$ ($\mu=1,2,3$) are the Pauli matrices
\footnote{Note that the representation of the gamma matrices in this paper is
different from the one we used in Ref.\cite{HSEK}. 
Our results are independent of the
representation we choose since these representations are unitary
equivalent to each other.}.

Now we consider time-reversal symmetry.
The time-reversal operator $\Theta$ is represented as
\be
\Theta=U\cdot K,
\ee
where $U$ is a unitary operator, and $K$ is the complex-conjugate operator.
The square of a time-reversal operator is either $+1$ or $-1$:
\be
\Theta_{\pm}^2=\pm 1.
\ee 
In the following, we focus on the time-reversal symmetry $\Theta_{+}$
with $\Theta_+^2=+1$.
For later use, it is convenient to choose $U$ in $\Theta_+$ as
$U=\Gamma_1\Gamma_4$.
Imposing the time-reversal invariance on $H({\bm k})$,
\be
\Theta_{+} H(-{\bm k}) \Theta_{+}^{-1} = H({\bm k}),\quad 
\Theta_{+}=\Gamma_1\Gamma_4\cdot K.
\label{timer_boson}
\ee
we obtain
\be
&&h_0({\bm k})=h_0(-{\bm k})^{*},\quad h_{(1,2)}({\bm k})=-h_{(1,2)}(-{\bm k})^{*},\quad h_{(3,4,5)}({\bm k})=h_{(3,4,5)}(-{\bm k})^{*},\nonumber\\
&&h_{(12,34,35,45)}({\bm k})=-h_{(12,34,35,45)}(-{\bm k})^{*},\quad
 h_{(13,14,15,23,24,25)}({\bm k})=h_{(13,14,15,23,24,25)}(-{\bm k})^{*}.~~~~~~~~
\label{rel1_boson}
\ee

In addition to the time-reversal invariance, we impose
pseudo-Hermiticity on $H({\bm k})$,
\be
\eta H({\bm k})^{\dag} \eta^{-1} = H({\bm k}), \quad \eta=i \Gamma_2\Gamma_1,
\label{pseudoHermiticity}
\ee
with Hermitian matrix $\eta$. $\eta$ is called the metric operator.
Here we have supposed that the metric operator $\eta$ anti-commutes with the
time-reversal operation,
\begin{eqnarray}
\{\Theta_+,\eta\}=0, 
\end{eqnarray}
which restricts the allowed form of $\eta$. 
From a proper unitary transformation and rescaling, we can always take
the basis in which  $\eta$ is given by $\eta=i\Gamma_2\Gamma_1$.
Equation (\ref{pseudoHermiticity}) leads to
\be
&&h_0({\bm k})=h_0({\bm k})^{*},\quad 
h_{(1,2)}({\bm k})=-h_{(1,2)}({\bm k})^{*},\quad
h_{(3,4,5)}({\bm k})=h_{(3,4,5)}({\bm k})^{*},\nonumber\\
&&h_{(12,34,35,45)}({\bm k})=h_{(12,34,35,45)}({\bm k})^{*},\quad
h_{(13,14,15,23,24,25)}({\bm k})=-h_{(13,14,15,23,24,25)}({\bm k})^{*}.~~~~~~~~
\label{rel2}
\ee
Combining (\ref{rel1_boson}) and (\ref{rel2}), $H({\bm k})$ is written as
\be
&&H({\bm k})=a_0({\bm k})+i\sum_{\mu=1,2} b_{\mu}({\bm k}) \Gamma_{\mu}
+\sum_{\mu=3,4,5} a_{\mu}({\bm k}) \Gamma_{\mu}\nonumber\\
&&+\sum_{\mu\nu=12,34,35,45} a_{\mu\nu}({\bm k}) \Gamma_{\mu\nu}
+i\sum_{\mu\nu=13,14,15,23,24,25} b_{\mu\nu}({\bm k}) \Gamma_{\mu\nu},
\label{hab_boson}
\ee
where $a_0({\bm k})$, $a_{\mu}({\bm k})$'s, $b_{\mu}({\bm k})$'s, $a_{\mu\nu}({\bm k})$'s, and $b_{\mu\nu}({\bm k})$'s 
are real functions of ${\bm k}$, and satisfy
\be
&&a_0({\bm k})=a_0(-{\bm k}),\quad 
b_{(1,2)}({\bm k})=b_{(1,2)}(-{\bm k}),\quad
a_{(3,4,5)}({\bm k})=a_{(3,4,5)}(-{\bm k}),\nonumber\\
&&a_{(12,34,35,45)}({\bm k})=-a_{(12,34,35,45)}(-{\bm k}),\quad
b_{(13,14,15,23,24,25)}({\bm k})=-b_{(13,14,15,23,24,25)}(-{\bm k}).~~~~~~~~
\label{even_odd2}
\ee

\subsection{$SO(3,2)$ Luttinger model and edge state}
\label{sec:spin_Chern}

In order to have real eigenenergies in non-Hermitian systems, 
$PT$ symmetry plays crucial roles as pointed out 
by Bender {\it et al.}\cite{Bender1,Bender2,Bender3}.
Following the arguments by Bender {\it et al.},
we impose the inversion symmetry on $H({\bm k})$ in (\ref{hab_boson}):
\be
 H(-{\bm k}) =H({\bm k}).
\label{inversion_boson}
\ee
From (\ref{hab_boson}), (\ref{even_odd2}), and (\ref{inversion_boson}), 
we obtain the following $SO(3,2)$ Hamiltonian \cite{HSEK}:
\be
H({\bm k})=a_0({\bm k})+i\sum_{\mu=1,2} b_{\mu}({\bm k}) \Gamma_{\mu}
+\sum_{\mu=3,4,5} a_{\mu}({\bm k}) \Gamma_{\mu},
\label{so32_Luttinger_boson}
\ee
with real even functions $a_0({\bm k})$, $a_{\mu}({\bm k})$'s, and $b_{\mu}({\bm k})$'s. 
The eigenenergies are 
\be
E_{\pm}(\bm k)=a_0(\bm k)\pm \sqrt{a_3(\bm k)^2+a_4(\bm k)^2+a_5(\bm k)^2
-b_1(\bm k)^2-b_2(\bm k)^2},
\ee
each of which is doubly degenerate.
As is expected from $PT$ symmetry, these eigenenergies are real if $a_3({\bm k})^2+a_4({\bm
 k})^2+a_5({\bm k})^2>b_1({\bm k})^2+b_2({\bm k})^2$.

To realize such an $SO(3,2)$ model,
we generalize the $SO(5)$ Luttinger Hamiltonian\cite{SM04,QiWuZhang} on
the square lattice into a non-Hermitian form: 
\be
H(\bm k)=\epsilon(\bm k)+V\sum_{a=3,4,5}d_{a}(\bm k)\Gamma_{a}
+i V\sum_{a=1,2}d_{a}\Gamma_{a},
\label{so32_boson}
\ee
where 
\be
&&\epsilon(\bm k)=t(-2\cos k_x-2\cos k_y +e_s),\quad
d_3(\bm k)=-\sqrt{3}\sin k_x \sin k_y, \nonumber\\
&&d_4(\bm k)=\sqrt{3}(\cos k_x - \cos k_y),\quad
d_5(\bm k)=2-e_s-\cos k_x - \cos k_y,
\label{dexp}
\ee
with ${\bm k}=(k_x,k_y)$.
The imaginary unit $i$ in the last term in the right hand side of
(\ref{so32_boson}) is absent in the original $SO(5)$ Luttinger Hamiltonian. 
For simplicity, we take $d_1$ and $d_2$ to be real constants. 
Here, $e_s\equiv \langle k_z^2 \rangle $
where $\langle\quad \rangle$ represents the expectation value in the lowest band\footnote{The model (\ref{so32_boson}) with $d_1=d_2=0$ corresponds to the Luttinger Hamiltonian in a symmetric quantum well along the $z$ direction.
When the quantum well is narrow enough, we could replace $k_z$-dependent
terms $\cos k_z \sim k_z^2$ with the expectation value 
$e_s\equiv \langle k_z^2 \rangle$
by eigenstates corresponding to the lowest 2D band 
(see Ref.\cite{QiWuZhang}).}.
The eigenenergies are 
\be
E_{\pm}(\bm k)=\epsilon(\bm k)\pm V\sqrt{d_3(\bm k)^2+d_4(\bm k)^2+d_5(\bm k)^2
-d_1^2-d_2^2},
\ee
each of which is doubly degenerate.

Now we examine edge states of this model.
We put the system on a cylinder with the periodic boundary 
condition in the $y$ direction and open boundary condition 
in the $x$ direction, and study the quasiparticle spectrum numerically.

\begin{figure}
\begin{center}
\includegraphics[width=6.0cm,clip]
{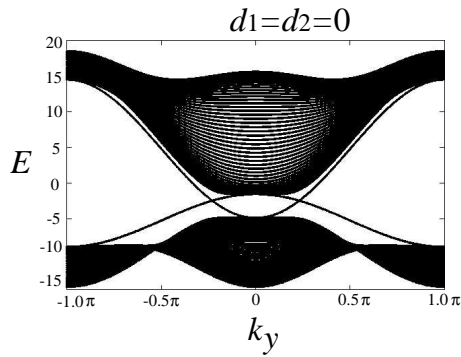}
\caption{The energy bands of the $SO(3,2)$ Luttinger Hamiltonian
 (\ref{so32_boson}) with open boundary
 condition in the $x$-direction for $t=1.0$, 
$V=4.0$, $e_s=0.5$, and $d_1=d_2=0$ (Hermitian case).
Here $k_y$ is the momentum in the $y$-direction.
Two gapless helical edge states appear on each edge.
}
\label{fig:edge_square}
\end{center}
\end{figure}

First, we illustrate the quasiparticle spectrum for the Hermitian case
with $d_1=d_2=0$.
The energy bands in this case are shown 
in Fig. \ref{fig:edge_square}.
In the gap of the bulk bands, there exist doubly degenerate edge bands.
The existence of the gapless edge bands is explained by so-called 
the spin Chern number \cite{QiWuZhang}.
When $d_1=d_2=0$, in addition to the 
$\Theta_{+}^2=+1$ time-reversal symmetry (\ref{timer_boson}),
the Hamiltonian (\ref{so32_boson}) has the following 
$\Theta_{-}^2=-1$ time-reversal symmetry:
\be
\Theta_{-} H(-{\bm k}) \Theta_{-}^{-1}=H({\bm k}), \quad 
\Theta_{-}=\Gamma_2\Gamma_4\cdot K.
\label{timer_fermion}
\ee
Moreover, since $H({\bm k})=H^{\dagger}({\bm k})$ for $d_1=d_2=0$, 
the pseudo-Hermiticity (\ref{pseudoHermiticity}) reduces to
\be
[H({\bm k}),\eta]=0.
\ee
Thus in this special case, $\eta$ becomes a conserved quantity. Indeed, the
operator $\eta$ is identified with the pseudo-spin operator $S_z$ in
the Luttinger model \cite{QiWuZhang} whose eigenvalue is either $+1$ or $-1$. 
In Ref.\cite{QiWuZhang}, the spin Chern number $C_s$ was defined by using
the pseudo-spin $S_z$ and the time-reversal symmetry $\Theta_-$, 
and it was shown that $C_s=2$ if $0<e_s<4$.
Therefore, the existence of the gapless edge bands is ensured by the
spin-Chern number.

\begin{figure}
\begin{center}
\includegraphics[width=8.5cm,clip]
{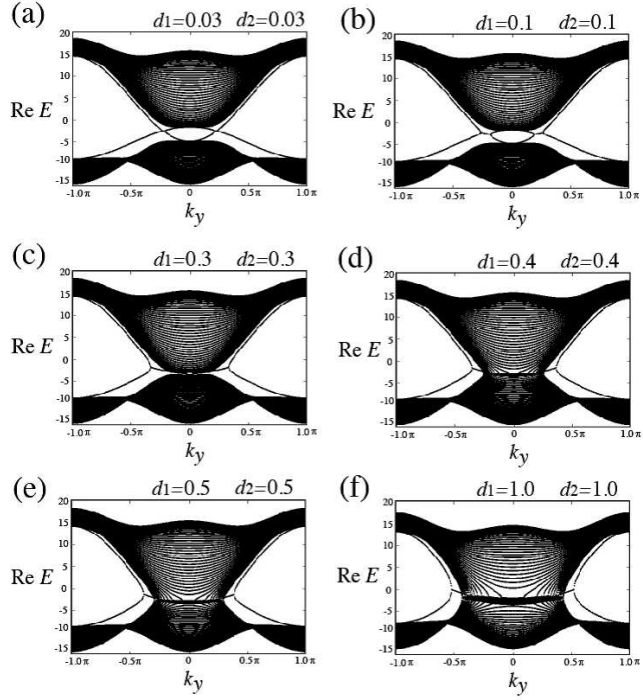}
\caption{The real part of the energy bands of the $SO(3,2)$ Luttinger
 Hamiltonian (\ref{so32_boson}) with the open boundary condition in the
 $x$-direction. 
We plot the results for $t=1.0$, $V=4.0$, $e_s=0.5$, and various
 values of the non-Hermitian parameters
 $(d_1,d_2)$. Here $k_y$ is the momentum in the $y$-direction.
For weak non-Hermiticity [(a) and (b)], two gapless helical edge modes
 survive. When the bulk gap in the real part of the energy closes [(c)-(d)], the
 topological number $C_{\rm TRI}$ is not well defined. In this parameter
 region, only remnants of the edge states are observed.} 
\label{fig:boson_square1}
\end{center}
\end{figure}

\begin{figure}
\begin{center}
\includegraphics[width=8.5cm,clip]
{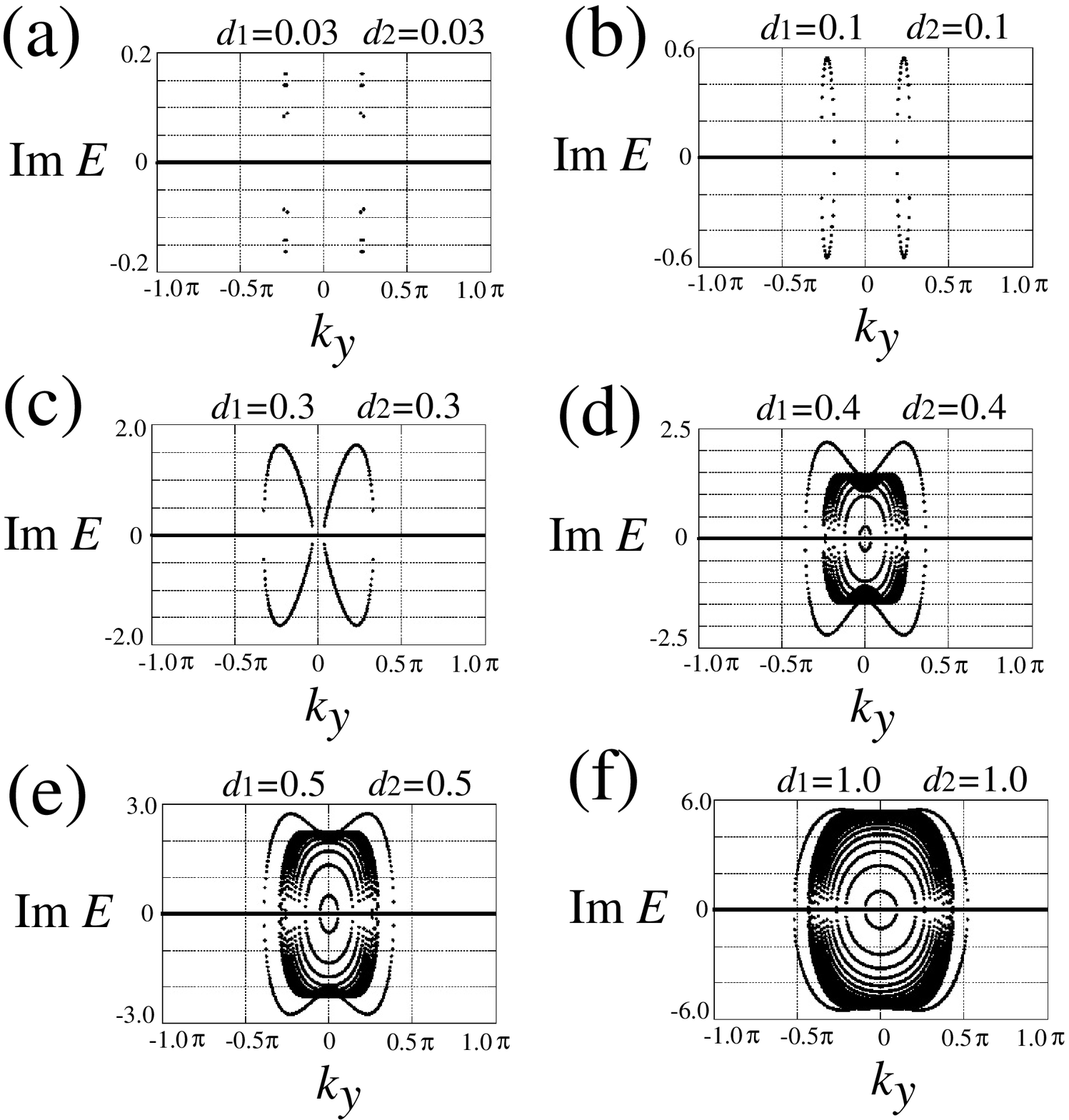}
\caption{
The imaginary part of the energy bands of the $SO(3,2)$ Luttinger
 Hamiltonian (\ref{so32_boson}) with the open boundary condition in the
 $x$-direction. 
We plot the results for $t=1.0$, $V=4.0$, $e_s=0.5$, and various
 values of the non-Hermitian parameters
 $(d_1,d_2)$. Here $k_y$ is the momentum in the $y$-direction.
For weak non-Hermiticity [(a) and (b)], there is no imaginary part in
 the bulk bands. Only the edge states with ${\rm Re}E=0$
 [Fig.\ref{fig:boson_square1} (a) and (b)] support the imaginary part in
 their energy.  
When the bulk gap in the real part of the energy closes
 [Fig.\ref{fig:boson_square1}(c)-(d)], the bulk states also support the
 imaginary part as well. 
}
\label{fig:boson_square2}
\end{center}
\end{figure}

Let us now examine the non-Hermitian case.
We show the real part and the imaginary one of the energy bands 
as functions of $k_y$ for various $(d_1,d_2)$'s 
in Figs. \ref{fig:boson_square1} and \ref{fig:boson_square2}, respectively.
Interestingly, the gapless edge bands 
persist in the real part even in the presence of the non-Hermiticity
$(d_1, d_2)$ [Figs.\ref{fig:boson_square1} (a) and (b)], 
although they have a small imaginary part at the same time
[Figs.\ref{fig:boson_square2} (a) and (b)]. 
Since the non-Hermiticity breaks both of the time-reversal
invariance for $\Theta_-$ and the rotational invariance for $S_z$,
such gapless modes cannot be explained by the spin Chern number.
The topological origin of the gapless edge mode will be discussed in the
next section.
If we further increase $(d_1,d_2)$, the bulk gap closes,   
and only remnants of the edge states are observed
[Figs.\ref{fig:boson_square1} (c)-(f)].


\subsection{Time-reversal invariant Chern number}
\label{sec:spin_Chern_so32}

In the previous section, we showed that there exist gapless edge states in
the $SO(3,2)$ non-Hermitian model.
In this section, we will discuss the topological origin of these gapless
edge states.

\subsubsection{Generalized Kramers theorem}

First of all, we examine general properties
of eigenstates for non-Hermitian Hamiltonian.
Let us denote the right eigenvectors and the left eigenvectors
of a non-Hermitian Hamiltonian $H({\bm k})$ as $\ket{u_n({\bm k})}$ and
$|u_n({\bm k}) \rangle\!\rangle$,  respectively:
\be
 H({\bm k})\ket{u_n({\bm k})}=E_n({\bm k})\ket{u_n({\bm k})},\quad
 H({\bm k})^{\dag} |u_n({\bm k}) \rangle\!\rangle
=E_n({\bm k})^*  |u_n({\bm k}) \rangle\!\rangle,
\label{eq:eigen}
\ee
where $\ket{u_n({\bm k})}$ and $ |u_n({\bm k}) \rangle\!\rangle$ are
normalized as
\be
 \langle\!\langle u_m({\bm k})|u_n({\bm k})\rangle
=\langle u_m({\bm k})|u_n({\bm k}) \rangle\!\rangle =\delta_{mn}.
\label{normalize_NH}
\ee
The eigenstates $|u_n({\bm k})\rangle$ and $|u_n({\bm k})
\rangle\!\rangle$ satisfying (\ref{normalize_NH}) are called as the
bi-orthonormal basis. 

Now suppose that $H({\bm k})$ is pseudo-Hermitian:
\begin{eqnarray}
H({\bm k})^{\dagger}=\eta H({\bm k}) \eta^{-1}. 
\end{eqnarray}
From the pseudo-Hermiticity, the first equation of (\ref{eq:eigen}) is
rewritten as 
\begin{eqnarray}
H({\bm k})^{\dagger}\eta|u_n({\bm k})\rangle
=E_n({\bm k}) \eta|u_n({\bm k})\rangle.
\end{eqnarray}
Therefore, $\eta|u_n({\bm k})\rangle$ is an eigenstate of $H({\bm
k})^{\dagger}$ and it can be expanded as
\begin{eqnarray}
\eta|u_n({\bm k})\rangle=\sum_m|u_m({\bm k})\rangle\!\rangle c_{mn}({\bm
 k}). 
\end{eqnarray}
Applying $\langle u_m({\bm k})|$ from the left, we obtain
\begin{eqnarray}
c_{mn}({\bm k})
=\langle u_m({\bm k})|\eta|u_n({\bm k})\rangle. 
\end{eqnarray}
Since $\eta$ is Hermitian, $c_{mn}({\bm k})$ is also Hermitian with
respect to the index $m$ and $n$. Thus it can be diagonalized by a
unitary matrix $G$:
\begin{eqnarray}
\sum_{lk}G_{ml}({\bm k})^{\dagger}c_{lk}({\bm k})G_{kn}({\bm
 k})=\lambda_m({\bm k})\delta_{mn},  
\end{eqnarray}
with real $\lambda_n({\bm k})$. The eigenvalue $\lambda_n({\bm k})$ is
not zero since $c_{mn}({\bm k})$ is invertible.
Taking the following new bi-orthonormal basis
\begin{eqnarray}
|\phi_n({\bm k})\rangle=\sum_m|u_m({\bm k})\rangle
 G_{mn}({\bm k})/\sqrt{|\lambda_n({\bm k})|},
\quad
|\phi_n({\bm k})\rangle\!\rangle=\sum_m|u_m({\bm k})\rangle\!\rangle
 G_{mn}({\bm k}) \sqrt{|\lambda_n({\bm k})|},
\label{eq:newbasis}
\end{eqnarray}
with $\langle \phi_m({\bm k})|\phi_n({\bm
k})\rangle\!\rangle=\langle\!\langle \phi_m({\bm k})|\phi_n({\bm
k})\rangle=\delta_{mn}$, 
we have
\begin{eqnarray}
\eta|\phi_n({\bm k})\rangle={\rm sgn}(\lambda_n({\bm k}))|\phi_n({\bm
 k})\rangle\!\rangle. 
\end{eqnarray}
Because of the continuity of the wave function, ${\rm
sgn}(\lambda_n({\bm k}))$ does not change in the whole region of the
momentum space.
Thus, the states in the new basis are classified into two, {\it i.e.}
states with 
\begin{eqnarray}
\eta|\phi_n({\bm k})\rangle=|\phi_n({\bm
 k})\rangle\!\rangle, 
\label{eq:eta+}
\end{eqnarray}
and those with
\begin{eqnarray}
\eta|\phi_n({\bm k})\rangle=-|\phi_n({\bm
 k})\rangle\!\rangle.
\label{eq:eta-}
\end{eqnarray}
Here we should note that the new bases $|\phi_n({\bm k})\rangle$ are no
longer eigenstates of $H({\bm k})$ unless $E_n({\bm k})$ is real.
Indeed, the right hand side of (\ref{eq:newbasis}) mixes the eigenstates
with $E_n({\bm k})$ and those with $E_n({\bm k})^*$.
However, the mixed states have a common real part of the energy. 
Therefore, even if the bulk energy $E_n({\bm k})$ is not real, 
the structure of the real part of the eigenenergy remains the same 
in this basis. 

If $H({\bm k})$ is invariant under the time-reversal symmetry
$\Theta_+$ with $\Theta_+^2=+1$ and $\{\Theta_+, \eta\}=0$, 
we have an additional structure, {\it i.e.} the generalized Kramers
degeneracy \cite{HSEK}:
Let $|\phi_n^{(+)}({\bm k})\rangle$
be a wavefunction satisfying (\ref{eq:eta+}),
\begin{eqnarray}
\eta|\phi_n^{(+)}({\bm k})\rangle=|\phi_n^{(+)}({\bm
 k})\rangle\!\rangle,
\end{eqnarray}
with
\begin{eqnarray}
\langle \phi_m^{(+)}({\bm k})|\phi_n^{(+)}({\bm k})\rangle\!\rangle
=\langle\!\langle \phi_m^{(+)}({\bm k})|\phi_n^{(+)}({\bm k})\rangle
=\delta_{mn}. 
\end{eqnarray}
The generalized Kramers partner $|\phi_n^{(-)}({\bm k})\rangle$ is given
by
\begin{eqnarray}
|\phi_n^{(-)}({\bm k})\rangle 
=e^{i\theta_n({\bm k})}\Theta\eta^{-1}|\phi_n^{(+)}(-{\bm k})\rangle\!\rangle,
\end{eqnarray}
with a phase factor $\theta_n({\bm k})$, and 
the corresponding left state is
\begin{eqnarray}
|\phi_n^{(-)}({\bm k})\rangle\!\rangle 
=e^{i\theta_n({\bm k})}\Theta\eta^{-1}|\phi_n^{(+)}(-{\bm k})\rangle,
\end{eqnarray} 
with
\begin{eqnarray}
\langle \phi_m^{(-)}({\bm k})|\phi_n^{(-)}({\bm k})\rangle\!\rangle
=\langle\!\langle \phi_m^{(-)}({\bm k})|\phi_n^{(-)}({\bm k})\rangle
=\delta_{mn}. 
\end{eqnarray}
The generalized Kramers pairs $|\phi_n^{(\pm)}({\bm
k})\rangle$ are independent of each other \cite{HSEK}. 
Actually, it is found easily that
at time-reversal invariant momenta ${\bm k}=\bar{\Gamma}_{i}$ 
satisfying $-\bar{\Gamma}_{i}=\bar{\Gamma}_{i}+{\bm G}$ 
with a reciprocal vector ${\bm G}$,
these generalized Kramers pairs are orthogonal to each other:
\begin{eqnarray}
\langle\!\langle \phi_n^{(+)}(\bar{\Gamma}_i)
|\phi_n^{(-)}(\bar{\Gamma}_i)\rangle=0, 
\end{eqnarray}
since we have
\begin{eqnarray}
\langle\!\langle \phi_n^{(+)}(\bar{\Gamma}_i)
|\phi_n^{(-)}(\bar{\Gamma}_i)\rangle&=& 
e^{i\theta_n(\bar{\Gamma}_i)}
\langle\!\langle \phi_n^{(+)}(\bar{\Gamma}_i)
|\Theta \eta^{-1} |\phi_n^{(+)}(\bar{\Gamma}_i)\rangle\!\rangle
\nonumber\\
&=&e^{i\theta_n(\bar{\Gamma}_i)}
\langle\!\langle \Theta^2 \eta^{-1}\phi_n^{(+)}(\bar{\Gamma}_i)
|\Theta \phi_n^{(+)}(\bar{\Gamma}_i)\rangle\!\rangle 
\nonumber\\
&=&e^{i\theta_n(\bar{\Gamma}_i)}
\langle\!\langle \phi_n^{(+)}(\bar{\Gamma}_i)
|\eta^{-1}\Theta \phi_n^{(+)}(\bar{\Gamma}_i)\rangle\!\rangle 
\nonumber\\
&=&-e^{i\theta_n(\bar{\Gamma}_i)}
\langle\!\langle \phi_n^{(+)}(\bar{\Gamma}_i)
|\Theta \eta^{-1}\phi_n^{(+)}(\bar{\Gamma}_i)\rangle\!\rangle 
\nonumber\\
&=&
-\langle\!\langle \phi_n^{(+)}(\bar{\Gamma}_i)
|\phi_n^{(-)}(\bar{\Gamma}_i)\rangle.
\end{eqnarray}

We also find that the generalized Kramers partner 
$|\phi_n^{(-)}({\bm k})\rangle$ satisfies (\ref{eq:eta-}),
\begin{eqnarray}
\eta |\phi_n^{(-)}({\bm k})\rangle&=&-e^{i\theta_n({\bm k})}
\Theta|\phi_n^{(+)}(-{\bm k})\rangle\!\rangle 
\nonumber\\
&=&-e^{i\theta_n({\bm k})}\Theta \eta^{-1}|\phi_n^{(+)}(-{\bm k})\rangle
\nonumber\\
&=&-|\phi_n^{(-)}({\bm k})\rangle\!\rangle.
\end{eqnarray}
Namely, in the presence of the time-reversal symmetry with $\Theta_+^2=+1$
and $\{\eta, \Theta_+\}=0$, the states with (\ref{eq:eta+}) are paired
with those with (\ref{eq:eta-}). 
As is shown below, this structure enables us to introduce a non-trivial
Chern number even for the time-reversal invariant system. 

\subsubsection{Chern numbers for non-Hermitian systems}

Here we generalize the Chern number for non-Hermitian systems.
For non-Hermitian systems, the gauge potential $A_i({\bm k})$ is introduced
as follows \cite{NesterovCruz2008,KH10_1}, 
\be
A_i({\bm k})=i \sum_{{\rm Re}E_n({\bm k})<E}\langle\!\langle u_n({\bm k})|
\frac{\partial}{\partial k_i}|u_n({\bm k})\rangle,
\label{gauge_NH}
\ee
where $\ket{u_n({\bm k})}$ denotes the right eigenstate of $H({\bm k})$ and
$|u_n({\bm k})\rangle\!\rangle$ the
corresponding left eigenstate. 
Here we suppose that there is a band gap at $E$ in the
real part, so no momentum ${\bm k}$ satisfies
${\rm Re}E_n({\bm k})=E$.
The gauge potential $A_i({\bm k})$ is well defined in
the whole region of the momentum space under this assumption.
The Chern number $C$ is defined as
\begin{eqnarray}
C=\frac{1}{2\pi}\int\int_{\rm FBZ} dk_xdk_y F_{xy}({\bm k}),
\label{eq:chernnumber1}
\end{eqnarray}
where $F_{xy}({\bm k})$ is the field strength of the gauge potential
$A_i({\bm k})$,
\be
F_{xy}({\bm k})=\frac{\partial
A_y({\bm k})}{\partial k_x}-\frac{\partial
A_x({\bm k})}{\partial k_y}, 
\ee
and the integration is performed in the first Brillouin zone.
In a manner similar to Hermitian systems, one can show that $C$ is
quantized and takes only integer values:
The Chern number (\ref{eq:chernnumber1}) counts vorticities
of wavefunctions in the first Brillouin zone \cite{TKNN2}.
By the gauge transformations 
$\ket{ u_n'({\bm k})}=e^{i\alpha({\bm k})}\ket{u_n({\bm k})}$ and
$|u_n'({\bm k})\rangle\!\rangle=e^{i\alpha({\bm k})}|u_n({\bm k})\rangle\!\rangle$,
the gauge potential $A_i({\bm k})$ is transformed as
\be
A_i({\bm k})'= A_i({\bm k})-\frac{\partial \alpha({\bm k})}{\partial k_i},
\label{gaugetrans}
\ee
where the normalization condition (\ref{normalize_NH}) was used.
As was shown in Ref.\cite{TKNN2}, the Chern number $C$ reduces to the line integral of the second
term of (\ref{gaugetrans}) around the zeros of the wavefunction,
\be
\frac{1}{2\pi}\oint \frac{\partial \alpha({\bm k})}{\partial {\bm k}}\cdot d{\bm k},
\ee
which counts the vorticity of the wavefunction in the first Brillouin
zone\cite{TKNN2}.

When the system is time-reversal invariant, one can show that the Chern
number $C$ is always zero. Thus we cannot use the Chern number itself to
characterize topological phases for time-reversal invariant systems.
However, if the system is pseudo-Hermitian as well and the time-reversal
symmetry $\Theta_+$ satisfies $\Theta_+^2=+1$ and $\{\eta,
\Theta_+\}=0$, one can define a different Chern number which can be
non-trivial even in the presence of the time-reversal invariance. 
The key structure is
the generalized Kramers pairs explained in the previous section.
Under the above assumption for $\Theta_+$ and $\eta$, 
one can take the following basis,
\begin{eqnarray}
\eta|\phi_n^{(\pm)}({\bm k})\rangle=\pm |\phi_n^{(\pm )}({\bm
 k})\rangle\!\rangle,
\label{eq:eta+-}
\end{eqnarray}
with
\begin{eqnarray}
\langle \phi_m^{(\pm )}({\bm k})|\phi_n^{(\pm )}({\bm k})\rangle\!\rangle
=\langle\!\langle \phi_m^{(\pm )}({\bm k})|\phi_n^{(\pm)}({\bm k})\rangle
=\delta_{mn}. 
\end{eqnarray}
Here the states $|\phi_n^{(\pm)}({\bm k})\rangle$ form the generalized
Kramers pair,
\begin{eqnarray}
|\phi_n^{(-)}({\bm k})\rangle 
=e^{i\theta_n({\bm k})}\Theta\eta^{-1}|\phi_n^{(+)}(-{\bm k})\rangle\!\rangle,
\quad
|\phi_n^{(-)}({\bm k})\rangle\!\rangle 
=e^{i\theta_n({\bm k})}\Theta\eta^{-1}|\phi_n^{(+)}(-{\bm k})\rangle.
\end{eqnarray} 
Using the sign difference of the right hand side of (\ref{eq:eta+-}), we
can introduce the following two different gauge potentials
\begin{eqnarray}
A_i^{(\pm )}({\bm k})=i\sum_{{\rm Re}E_n({\bm k})<E}
\langle\!\langle \phi_n^{(\pm)}({\bm k})| 
\frac{\partial}{\partial k_i}|\phi_n^{(\pm)}({\bm k})\rangle,
\label{eq:gauge+-}
\end{eqnarray}
and the corresponding Chern numbers,
\begin{eqnarray}
C^{(\pm)}=\frac{1}{2\pi}\int\int_{\rm FBZ} dk_xdk_y F^{(\pm)}_{xy}({\bm k}),
\label{eq:chernnumber}
\end{eqnarray}
with
\be
F_{xy}^{(\pm)}({\bm k})=\frac{\partial
A_y^{(\pm)}({\bm k})}{\partial k_x}-\frac{\partial
A_x^{(\pm)}({\bm k})}{\partial k_y}.
\ee
As we mentioned in the previous section, if $E_n({\bm k})$ is not real,
the states $|\phi_n^{(\pm)}({\bm k})\rangle$ are not eigenstates of
$H({\bm k})$ in general. 
However, for the real part of the eigenenergy, $|\phi_n^{(\pm)}({\bm
k})\rangle$ has the same structure as the eigenstate of $H({\bm k})$.
Thus the summation in (\ref{eq:gauge+-}) is well defined.
It is also found that 
\begin{eqnarray}
C=C^{(+)}+C^{(-)}. 
\end{eqnarray}
Under the time-reversal transformation, these Chern numbers transform as
$C^{(\pm)}\rightarrow -C^{(\mp)} $. 
Thus the time-reversal invariance implies that $C=C^{(+)}+C^{(-)}=0$.
On the other hand, the time-reversal invariant combination
\begin{eqnarray}
C_{\rm TRI}=\frac{C^{(+)}-C^{(-)}}{2}, 
\end{eqnarray}
can be non-zero. We call $C_{\rm TRI}$ as time-reversal invariant Chern
number.
In Sec. \ref{sec:VC3} and Sec. \ref{sec:so32}, we will see that the
time-reversal invariant Chern number characterizes the
gapless edge state for a class of non-Hermitian systems.

\subsubsection{Application to $SO(3,2)$ model}
\label{sec:VC3}

Let us now examine $C_{\rm TRI}$ for the $SO(3,2)$ model
(\ref{so32_Luttinger_boson}).
Below, we suppose that we have a gap around $E_{-}(\bm k)$ and $E_{+}(\bm k)$,
and that the system is half-filling.
In this model, two right eigenstates $\ket{u^{(+)}_-({\bm k})}$ and 
$\ket{u_-^{(-)}({\bm k})}$ with the negative eigenenergy $E_{-}(\bm k)$ are
\be
\ket{u_-^{(+)}{({\bm k})}}=\frac{1}{\sqrt{2E(E-a_5)}}
\left(
\begin{array}{c} 
    a_5-E \\ 
   a_3+ia_4  \\
ib_1-b_2\\
0 \\
\end{array} \right),\quad
\ket{u_-^{(-)}({\bm k})}=\frac{1}{\sqrt{2E(E-a_5)}}
\left(
\begin{array}{c} 
    0 \\ 
  ib_1+b_2  \\
 -a_3+i a_4\\
a_5-E \\
\end{array} \right),~~~~
\label{eigenstates}
\ee
where $E\equiv\sqrt{a_3^2+a_4^2+a_5^2-b_1^2-b_2^2}$.
They form the generalized Kramers pair.

Here we present explicit calculations of the time-reversal invariant
Chern number
for $SO(3,2)$ Hamiltonians (\ref{so32_Luttinger_boson}).
Substituting (\ref{eigenstates}) into (\ref{eq:gauge+-}),
we obtain
\be
A_{i}^{(+)}({\bm k})=\frac{i}{2E(E-a_5)}
\left[(E-a_5)\frac{\partial}{\partial k_i} (E-a_5) 
+ (a_3-i a_4)\frac{\partial}{\partial k_i} (a_3+i a_4) \right. \nonumber\\
\left. + (i b_1+ b_2)\frac{\partial}{\partial k_i} (i b_1- b_2)\right]
+i\frac{\partial}{\partial k_i}
\left[\ln\left(\frac{1}{\sqrt{2E(E-a_5)}}\right)\right],
\nonumber\\
A_{i}^{(-)}({\bm k})=\frac{i}{2E(E-a_5)}
\left[(E-a_5)\frac{\partial}{\partial k_i} (E-a_5) 
+ (a_3+i a_4)\frac{\partial}{\partial k_i} (a_3-i a_4) \right. \nonumber\\
\left. + (i b_1- b_2)\frac{\partial}{\partial k_i} (i b_1+ b_2)\right]
+i\frac{\partial}{\partial k_i}
\left[\ln\left(\frac{1}{\sqrt{2E(E-a_5)}}\right)\right].
\label{gaugeexplicit}
\ee
From (\ref{gaugeexplicit}), we find
\be
A_{i}^{(-)}({\bm k})=[A_{i}^{(+)}(-{\bm k})]^{*},
\label{gaugerelation}
\ee
from which we can show $C=C^{(+)}+C^{(-)}=0$ explicitly.

To evaluate the time-reversal invariant Chern number, we adiabatically
deform the Hamiltonian of the system without gap closing in the real
part of the bulk energy.
This process does not change the time-reversal invariant Chern number. 
In particular, to calculate the time-reversal invariant Chern number for
weak non-Hermiticity in Figs.\ref{fig:boson_square1} (a) and (b), we
decrease the
non-Hermiticity $(d_1, d_2)$ adiabatically as $d_i\rightarrow 0$ $(i=1,2)$.
In this particular limit, we find that the time-reversal invariant Chern
number $C_{\rm TRI}$ coincides with the spin Chern number $C_s$ in
Ref.\cite{QiWuZhang}. 
Therefore, from the adiabatic continuity, we obtain
\begin{eqnarray}
C_{\rm TRI}=2, 
\end{eqnarray}
for the model in Fig.\ref{fig:boson_square1}.
This means that the existence of the gapless edge states 
in Fig.\ref{fig:boson_square1}
is ensured by the non-zero value of $C_{\rm TRI}$. 
Here we should emphasize that the spin Chern number itself is not
well-defined once the non-Hermiticity $(d_1, d_2)$ is turned on.
On the other hand, the time-reversal invariant Chern number $C_{\rm
TRI}$ is well defined even in the presence of the non-Hermiticity.

\section{Non-Hermitian Kane-Mele model}
\label{sec:so32}

\begin{figure}
 \begin{center}
  \includegraphics[width=7.0cm,clip]{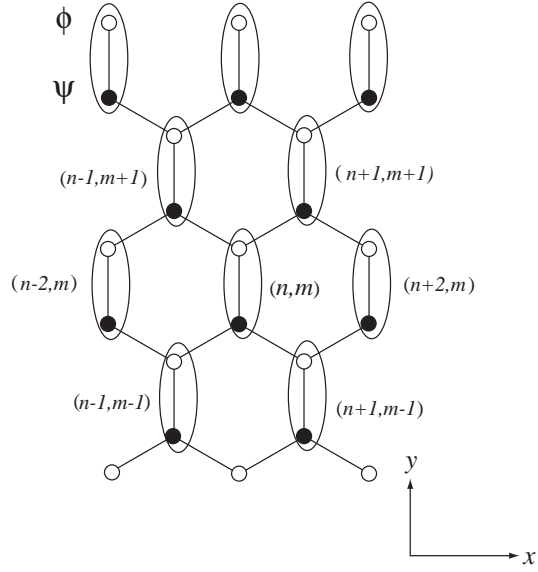}
\caption{\label{fig:honeycomb_lattice_spin}
 The honeycomb lattice.}
\end{center}
\end{figure} 
\begin{figure}
 \begin{center}
  \includegraphics[width=8.0cm,clip]{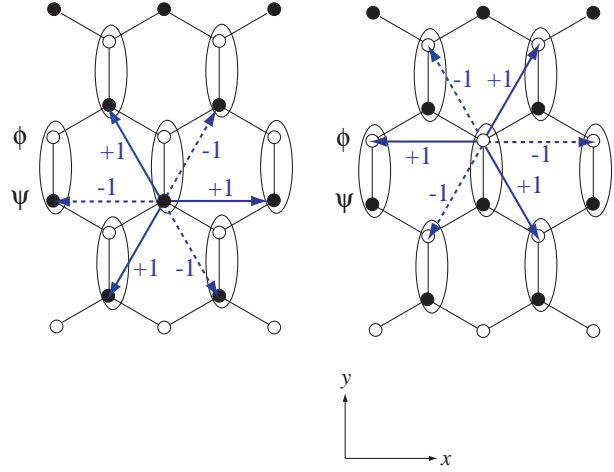}
\caption{\label{fig:honeycomb_lattice_nu}
 The sign of $\nu_{ij}$. }
\end{center}
\end{figure} 
\begin{figure}
 \begin{center}
  \includegraphics[width=8.0cm,clip]{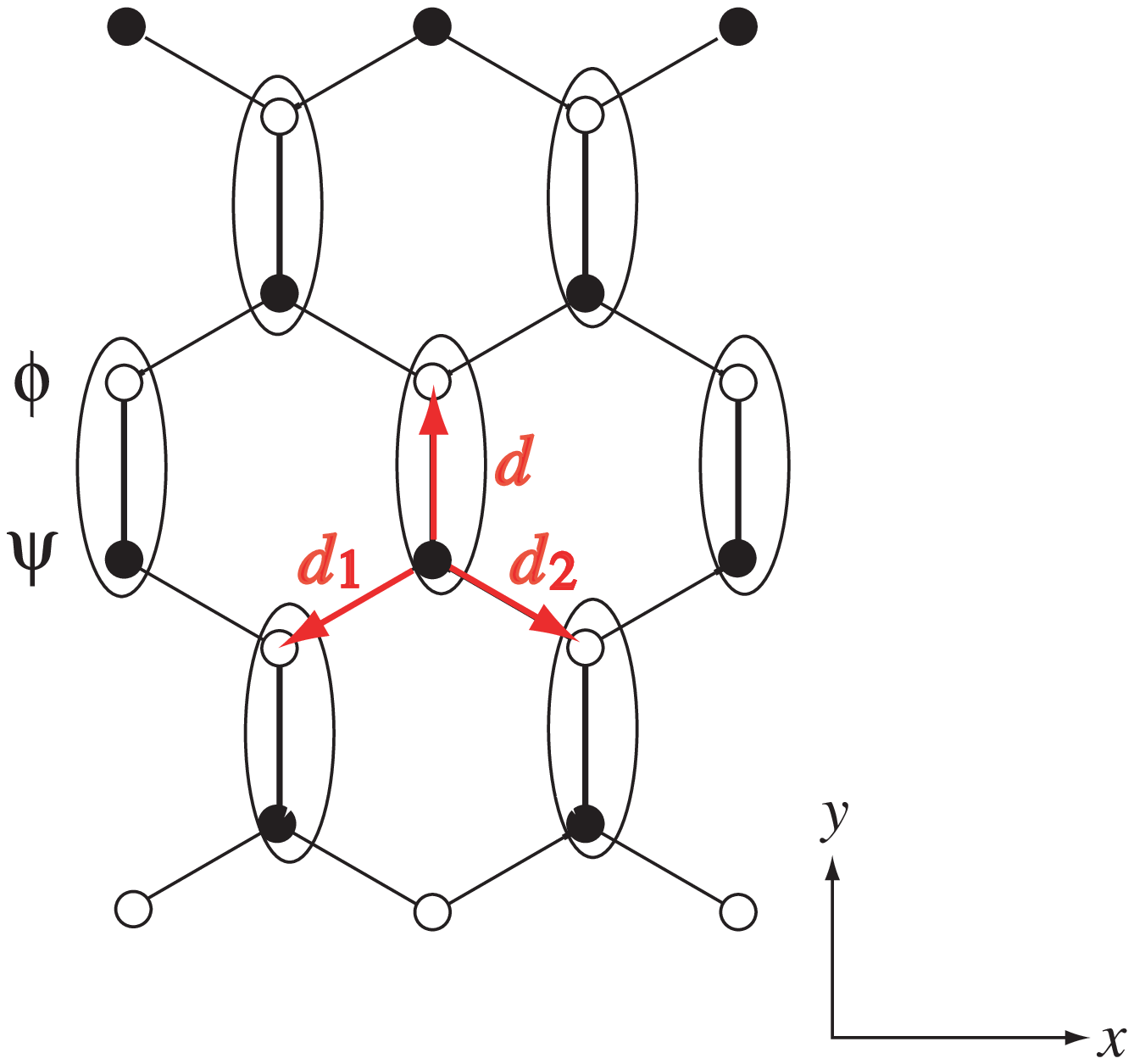}
\caption{\label{fig:honeycomb_lattice_d}
 ${\bm d}$, ${\bm d}_1$, and ${\bm d}_2$. }
\end{center}
\end{figure} 
\begin{figure}
 \begin{center}
  \includegraphics[width=6.0cm,clip]{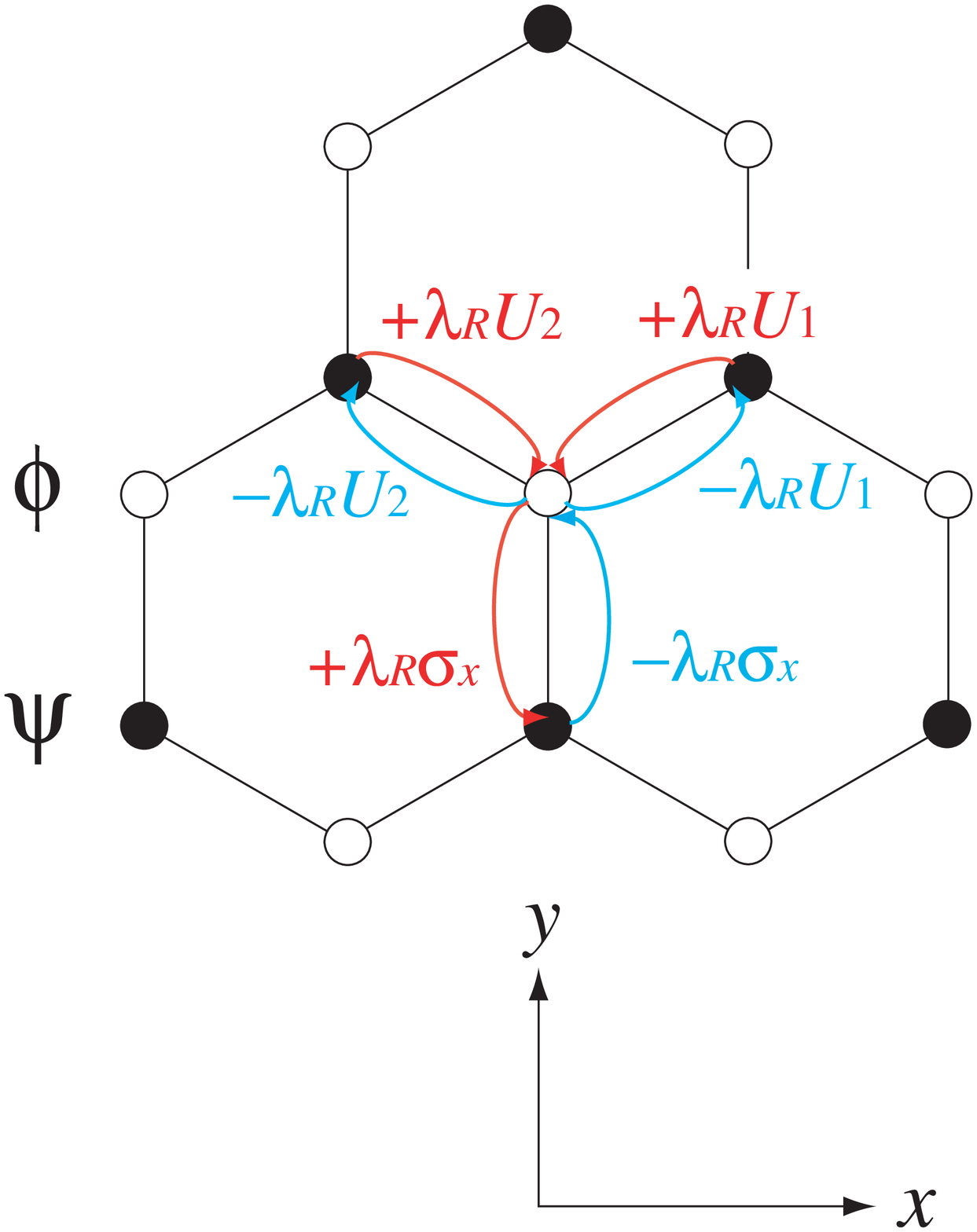}
\caption{\label{fig:Rashba}
The nearest-neighbor hopping integrals of
{\it imaginary} Rashba terms.
The hopping integrals of each bond are asymmetric.
[$U_1=\frac{1}{2}\sigma_x-\frac{\sqrt{3}}{2}\sigma_y$
and $U_2=\frac{1}{2}\sigma_x+\frac{\sqrt{3}}{2}\sigma_y$.]
}
\end{center}
\end{figure} 

So far, we argued systems with imaginary on-site potentials. 
In this section, we consider a system in which
the non-Hermiticity is caused by asymmetric hopping integrals. 

Consider the following non-Hermitian version of Kane-Mele
model\cite{KM05,KM05_2}:
\be
H=H_K+H_V+H_{SO}+\widetilde{H}_R,
\label{so32_Hamiltonian}
\ee
where
\be
&&H_K=t\sum_{\langle i,j\rangle} (c_i^{\dag} c_j+{\rm H.c.}),\quad
H_V=\lambda_V\sum_i \xi_i c_i^{\dag}c_i,\quad
H_{\rm SO}=i \lambda_{{\rm SO}}\sum_{\langle\!\langle i,j \rangle\!\rangle}
( \nu_{ij} c_i^{\dag} s^{z} c_j+{\rm H.c.}),\nonumber\\
&&\widetilde{H}_R=-\lambda_{{\rm R}}\sum_{\langle i,j \rangle} 
\left[c_i^{\dag}({\bm s}\times {\hat{\bm d}}_{ij})_z c_j+{\rm H.c.}\right].
\ee
The first term $H_K$ in (\ref{so32_Hamiltonian}) 
is the nearest-neighbor hopping.
The second term $H_V$ is sublattice potentials with 
 $\xi_i=+1$ ($-1$) for closed (open) circles in
 Fig.\ref{fig:honeycomb_lattice_spin}.
The third term $H_{\rm SO}$ represents the next-nearest-neighbor 
spin-orbit interaction
preserving the $z$ component $S_z$ of the spin, with $\nu_{ij}=\pm 1$.
(See Fig.\ref{fig:honeycomb_lattice_nu}.)
The last term $\widetilde{H}_R$ is the {\it imaginary} Rashba term 
which gives rise to non-Hermiticity.
Here, $\hat{{\bm d}}={\bm {\hat y}}$, $\hat{{\bm
d}}_1=-\frac{\sqrt{3}}{2} \hat{{\bm x}}-\frac{1}{2} \hat{{\bm y}}$, and
$\hat{{\bm d}}_2=+\frac{\sqrt{3}}{2} \hat{{\bm x}}-\frac{1}{2} \hat{{\bm
y}}$, are illustrated in Fig.\ref{fig:honeycomb_lattice_d}.

The {\it imaginary} Rashba term $\widetilde{H}_R$  
gives asymmetric nearest-neighbor hopping integrals. 
The term $\widetilde{H}_R$ is explicitly written as
\be
\widetilde{H}_R
&=&-\lambda_{{\rm R}}\sum_{\langle i,j\rangle} 
\left[c_i^{\dag}({\bm s}\times {\hat{\bm d}}_{ij})_z c_j +{\rm H.c.}\right]
\nonumber \\ 
&=&\lambda_{{\rm R}}\sum_{i\bullet} \left[- c_{i+{\bm d}}^{\dag}\sigma_x c_i
+c_{i+{\bm d}_1}^{\dag}U_1 c_i
+c_{i+{\bm d}_2}^{\dag}U_2 c_i
\right]
- \lambda_{{\rm R}} \sum_{i\circ} \left[-c_{i-{\bm d}}^{\dag}\sigma_x c_i
+c_{i-{\bm d}_1}^{\dag}U_1 c_i
+c_{i-{\bm d}_2}^{\dag}U_2 c_i
\right],
\label{Rashba_explicit}
\ee
where $\sum_{i\bullet}$ ($\sum_{i\circ}$) denotes the summation
over closed (open) circles in Fig.\ref{fig:honeycomb_lattice_spin},
$U_1\equiv\frac{1}{2}\sigma_x-\frac{\sqrt{3}}{2}\sigma_y$,
and $U_2\equiv\frac{1}{2}\sigma_x+\frac{\sqrt{3}}{2}\sigma_y$.
The hopping integrals are asymmetric for each bond as illustrated in
Fig.\ref{fig:Rashba}:
Let us respectively denote two eigenstates of $\sigma_x$,
$U_1$, and $U_2$
as $\ket{\pm}_x$, $\ket{\pm}_1$, and $\ket{\pm}_2$: 
\be
 \sigma_x \ket{\pm}_x=\pm \ket{\pm}_x,\quad
 U_1 \ket{\pm}_1=\pm \ket{\pm}_1,\quad
 U_2 \ket{\pm}_2=\pm \ket{\pm}_2.
\ee
They are explicitly given  by
\be
 \ket{\pm}_x=\frac{1}{\sqrt{2}}\left(
\begin{array}{c} 
    1 \\ 
   \pm 1  \\
\end{array} \right),\quad
 \ket{\pm}_1=\frac{1}{\sqrt{2}}\left(
\begin{array}{c} 
    e^{i\pi/3} \\ 
   \pm 1  \\
\end{array} \right),\quad
 \ket{\pm}_2=\frac{1}{\sqrt{2}}\left(
\begin{array}{c} 
    e^{-i\pi/3}  \\ 
   \pm 1  \\
\end{array} \right). 
\ee
For each eigenstate, the hopping integrals $t+\lambda_R\sigma_x$, $t+\lambda_R
U_1$, and $t+\lambda_R U_2$ read
\be
(t+\lambda_R\sigma_x)\ket{\pm}_x=(t\pm \lambda_R)\ket{\pm}_x,\quad
(t+\lambda_R U_{1(2)})\ket{\pm}_{1(2)}=(t\pm \lambda_R)\ket{\pm}_{1(2)},~~~~
\label{asymmetry_1}
\ee
and the hopping integrals 
$t-\lambda_R\sigma_x$, $t-\lambda_R U_1$, and $t-\lambda_R U_2$ yield
\be
(t-\lambda_R\sigma_x)\ket{\pm}_x=(t\mp \lambda_R)\ket{\pm}_x,\quad
(t-\lambda_R U_{1(2)})\ket{\pm}_{1(2)}=(t\mp \lambda_R)\ket{\pm}_{1(2)}.~~~~
\label{asymmetry_2}
\ee
The sign difference between (\ref{asymmetry_1}) and (\ref{asymmetry_2})
implies that the hopping integrals (for the basis
$(\ket{\pm}_x,\ket{\pm}_1,\ket{\pm}_2)$)
along the bonds parallel to $({\bm d}, {\bm d}_1, {\bm d}_2)$
are asymmetric: $(t+\lambda_R)$ in one direction and
 $(t-\lambda_R)$ in the other.

We also note that the 
{\it imaginary} Rashba spin-orbit interaction
can be regarded as an {\it imaginary} $SU(2)$ gauge potential  
[see Appendix \ref{sec:Appendix_SU2}].
In this sense, our model can be regarded as a non-Abelian generalization
of the Hatano-Nelson model in which an {\it imaginary} $U(1)$ vector
potential \cite{NH06,NH07,NH08} was considered.

Now we consider the Hamiltonian in the momentum space.
By performing the Fourier transformation with respect to
$(n, m)$ in Fig.\ref{fig:honeycomb_lattice_spin},
the Hamiltonian in the momentum space is obtained as
\be
H({\bm k})=\sum_{a=1,2} d_a({\bm k})\widetilde{\Gamma}_a
+\sum_{ab=12,15} d_{ab}({\bm k})\widetilde{\Gamma}_{ab}
+i \sum_{a=3,4} d_a({\bm k})\widetilde{\Gamma}_a
+i \sum_{ab=23,24} d_{ab}({\bm k})\widetilde{\Gamma}_{ab},~~~~
\label{KaneMele}
\ee
where 
$\widetilde{\Gamma}_{ab}=[\widetilde{\Gamma}_a,\widetilde{\Gamma}_b]/(2i)$,
and
\be
&&d_1=t(1+2\cos x\cos y),\quad
d_2=\lambda_V, \nonumber\\
&&d_3=\lambda_{\rm R} (1-\cos x\cos y),\quad 
d_4=-\sqrt{3}\lambda_{\rm R}\sin x\sin y,\quad\nonumber\\
&&d_{12}=-2t\cos x \sin y,\quad ~~~~~d_{15}=\lambda_{{\rm SO}}(2\sin 2x -4\sin x\cos y),
\nonumber\\ 
&&d_{23}=-\lambda_{\rm R}\cos x\sin y, \quad ~~~~
d_{24}=\sqrt{3}\lambda_{\rm R}\sin x \cos y,
\label{KaneMele_da_dab}
\ee
with $x=k_x a/2$ and  $y=\sqrt{3} k_y a/2$. 
Here, we have adopted the following gamma matrices:
\be
\widetilde{\Gamma}_{(1,2,3,4,5)}=(\sigma^x \otimes I,\sigma^z\otimes I,
\sigma^y\otimes \sigma^x, \sigma^y\otimes \sigma^y,\sigma^y\otimes \sigma^z).
\label{gamma_rashba}
\ee

The Hamiltonian (\ref{KaneMele}) possesses the pseudo-Hermiticity,
\be
\eta H({\bm k})^{\dag} \eta^{-1}=H({\bm k}),
\quad \eta=i\widetilde{\Gamma}_4\widetilde{\Gamma}_3,
\label{pseudo_KaneMele}
\ee
and time-reversal invariance  with $\widetilde{\Theta}_{+}^2=+1$,
\be
\widetilde{\Theta}_{+} H(-{\bm k}) \widetilde{\Theta}_{+}^{-1}=H({\bm k}), 
\quad \widetilde{\Theta}_{+}=i\widetilde{\Gamma}_5\widetilde{\Gamma}_4\cdot K. 
\label{boson_KaneMele}
\ee
These symmetries are simply understood by noticing that  
the Hamiltonian (\ref{KaneMele}) is a special case of (\ref{hab_boson}) 
by identifying  
$(\widetilde{\Gamma}_1,\widetilde{\Gamma}_2,\widetilde{\Gamma}_3,
\widetilde{\Gamma}_4,\widetilde{\Gamma}_5)$ in (\ref{gamma_rashba}) with $(\Gamma_3,\Gamma_5,\Gamma_2,\Gamma_1,\Gamma_4)$ 
in (\ref{gammaso32}). 
From the anticommutation relation $\{ \eta,\widetilde{\Theta}_{+} \}=0$,
the time-reversal invariant Chern number $C_{\rm TRI}$ can be introduced 
in a manner similar to the previous sections.
In the following, by using the time-reversal invariant  Chern number,
we argue the topological stability of 
the edge states of the Hamiltonian (\ref{so32_Hamiltonian}) 
with zigzag edges  [Fig.\ref{fig:honeycomb_lattice_zigzag}].

\begin{figure}
 \begin{center}
\includegraphics[width=6.0cm,clip]
{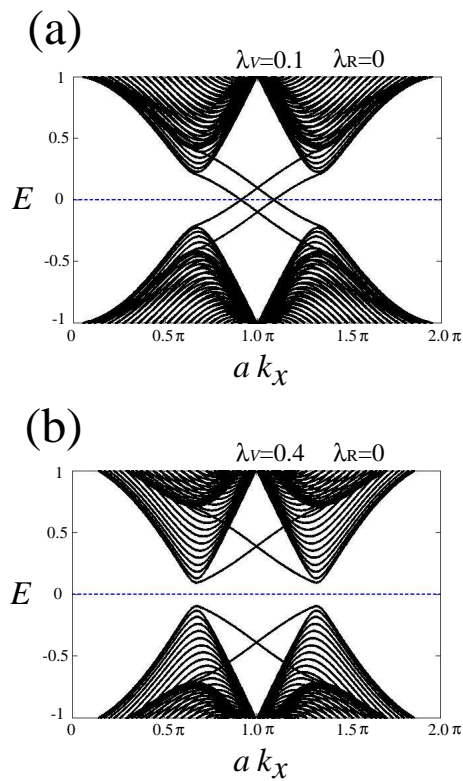}
\caption{The energy bands of the Kane-Mele model
  (\ref{so32_Hamiltonian}) with zigzag edges  along the $x$-direction
for $t=1.0$, $\lambda_{\rm SO}=0.06$, $\lambda_{\rm R}=0$ (Hermitian
  case), and (a) $\lambda_V=0.1$, (b)$\lambda_V=0.4$.
Here $a$ is the lattice constant, and $k_x$ the momentum in the
  $x$-direction. In (a), a gapless helical edge state appears on each
  edge, while in (b) no gapless edge state is obtained. 
}
\label{fig:so5}
\end{center}
\end{figure}

First, discuss the Hermitian case $\lambda_{{\rm R}}=0$
\cite{KM05,KM05_2}.
For $\lambda_{{\rm R}}=0$, in addition to the 
time-reversal symmetry $\widetilde{\Theta}_{+}$ with
$\widetilde{\Theta}_{+}^2=+1$ (\ref{boson_KaneMele}),
 the Hamiltonian (\ref{KaneMele}) has another time-reversal symmetry
$\widetilde{\Theta}_{-}$  with $\widetilde{\Theta}_{-}^2=-1$: 
\be
\widetilde{\Theta}_{-} H(-{\bm k}) \widetilde{\Theta}_{-}^{-1}=H({\bm k}), 
\quad \widetilde{\Theta}_{-}=\widetilde{\Gamma}_3\widetilde{\Gamma}_5\cdot K.
\ee
Moreover, for $\lambda_R=0$, the pseudo-Hermiticity (\ref{pseudo_KaneMele})
reduces to
\be
[H({\bm k}),\eta]=0.
\ee
Thus the metric operator $\eta$ becomes a conserved quantity
whose eigenvalue is either $+1$ or $-1$. 
By regarding $\eta$ as the $z$-component of the spin, the
spin Chern number $C_s$ can be defined \cite{Sheng}.
For $\lambda_V<3\sqrt{3}\lambda_{{\rm SO}}$, 
where the spin Chern number is non-zero, {\it i.e.} $C_s=1$, 
gapless edge modes appear as shown in Fig.\ref{fig:so5} (a).
On the other hand, for $\lambda_V>3\sqrt{3}\lambda_{{\rm SO}}$, 
the system is in the topologically trivial insulating phase with $C_s=0$.
In this phase, gapless edge modes do not appear as illustrated in
Fig.\ref{fig:so5} (b).

\begin{figure}
 \begin{center}
\includegraphics[width=6.0cm,clip]
{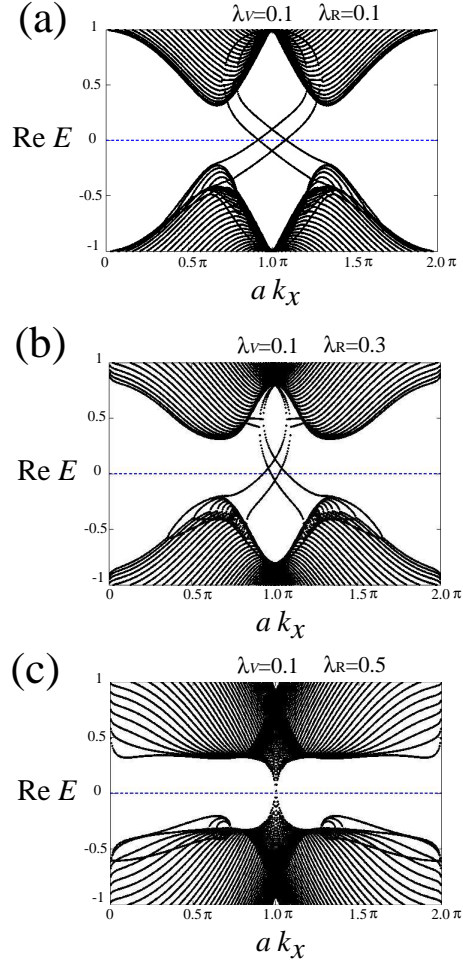}
\caption{The real part of the energy bands of the non-Hermitian
  Kane-Mele model (\ref{so32_Hamiltonian}) with zigzag edge along
  $x$-direction.
We plot the results for $t=1.0$, $\lambda_{\rm SO}=0.06$, 
$\lambda_V=0.1$, and various values of non-Hermitian parameter
  $\lambda_{{\rm R}}$. 
Here $a$ is the lattice constant, and $k_x$ is the momentum in the
  $x$-direction.
In (a) and (b), we have gapless edge states.
In (c), the bulk gap closes, and no gapless edge state exists. 
}
\label{fig:so32_5}
\end{center}
\end{figure}

\begin{figure}
 \begin{center}
\includegraphics[width=6.0cm,clip]
{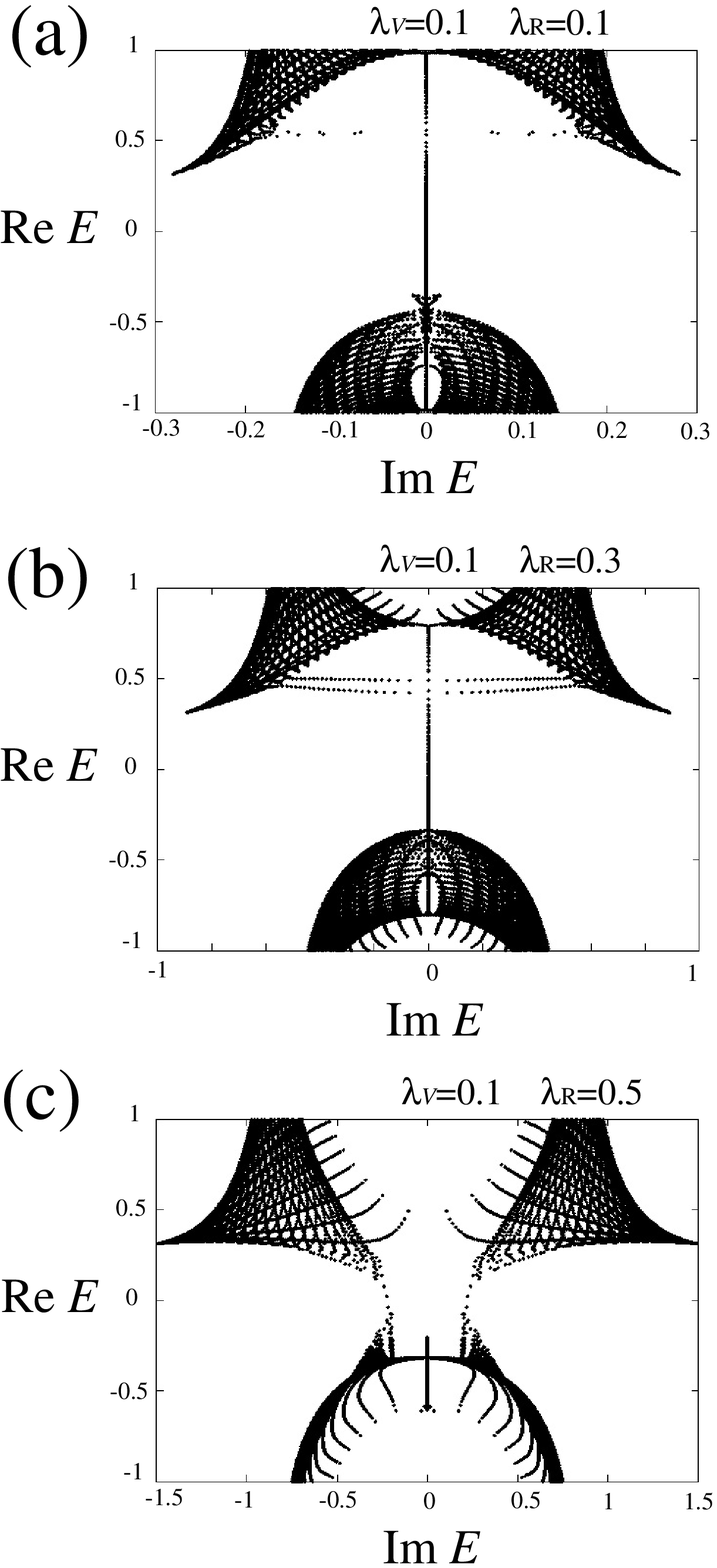}
\caption{The real part vs imaginary part of the energy bands
of non-Hermitian Kane-Mele model (\ref{so32_Hamiltonian}) with zigzag
  edges along the $x$-direction.
We plot the results for $t=1.0$, $\lambda_{\rm SO}=0.06$, 
$\lambda_V=0.1$, and various values of the non-Hermitian parameter 
$\lambda_{{\rm R}}$.
The data (a) and (b) indicate that the energy of the gapless edge
  states [Fig.\ref{fig:so32_5} (a) and (b)] is real.
On the other hand, the bulk state supports the imaginary part of the energy.  
}
\label{fig:so32_6}
\end{center}
\end{figure}

Now we include the non-Hermitian term $\lambda_{{\rm R}}$.
Once the non-Hermitian term is nonzero, $\eta$ is no longer conserved, 
thus the spin Chern number $C_s$ is not well defined.
Using the time-reversal invariant Chern number $C_{\rm
TRI}$, however, we can argue the topological stability of the gapless
edge modes.
Let us first consider the region $\lambda_V<3\sqrt{3}\lambda_{{\rm SO}}$
as shown in Figs.\ref{fig:so32_5} and \ref{fig:so32_6}.
When $\lambda_{{\rm R}}$ is small, the gapless edge modes that appear 
in the Hermitian case ($\lambda_{{\rm R}}=0$) still remain
[Figs.\ref{fig:so32_5} (a) and (b)].
These gapless edge states are topologically protected by
the time-reversal invariant Chern number $C_{\rm TRI}=1$:
For small $\lambda_{\rm R}$, the non-Hermitian Hamiltonian
(\ref{KaneMele})
can be adiabatically deformed into the Hermitian one ($\lambda_{{\rm R}}=0$)
without closing the bulk gap, and 
the time-reversal invariant Chern number $C_{\rm TRI}$
reduces to the spin Chern number $C_s$ in the Hermitian limit.
Thus, from the adiabatic continuity,  
we have $C_{\rm TRI}=C_s=1$. 
On the other hand, for sufficiently large $\lambda_{{\rm R}}$, the bulk
gap closes and $C_{\rm TRI}$ becomes ill-defined.  
Correspondingly, the gapless edge modes disappear.

\begin{figure}
 \begin{center}
\includegraphics[width=6.0cm,clip]
{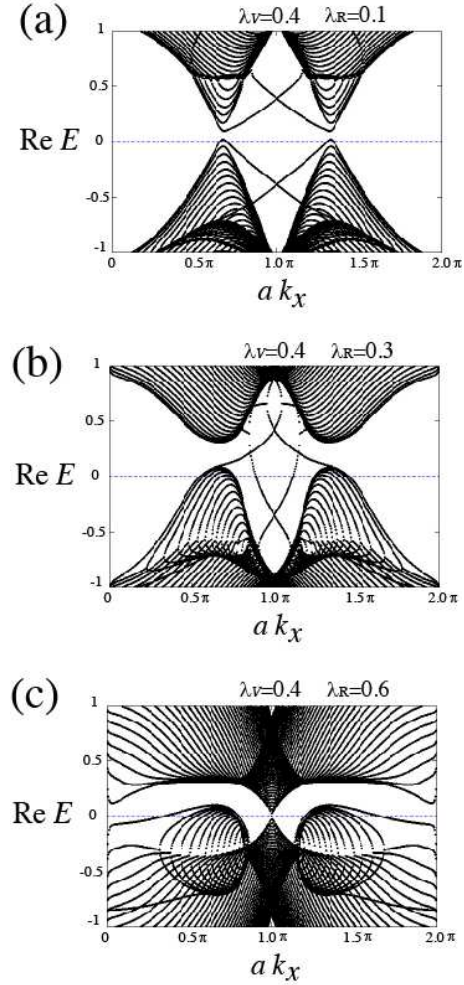}
\caption{The real part of the energy bands of the non-Hermitian
  Kane-Mele model (\ref{so32_Hamiltonian}) with zigzag edge along
  $x$-direction.
We plot the results for $t=1.0$, $\lambda_{\rm SO}=0.06$, 
$\lambda_V=0.4$, and various values of non-Hermitian parameter
  $\lambda_{{\rm R}}$. 
Here $a$ is the lattice constant, and $k_x$ is the momentum in the
  $x$-direction.
In (a) no gapless edge state exists, while in (b) topologically
  protected gapless edge states appear.
In (c), the bulk gap closes and the topologically protected gapless edge
  states disappear.
}
\label{fig:so32_7}
\end{center}
\end{figure}

\begin{figure}
 \begin{center}
\includegraphics[width=6.0cm,clip]
{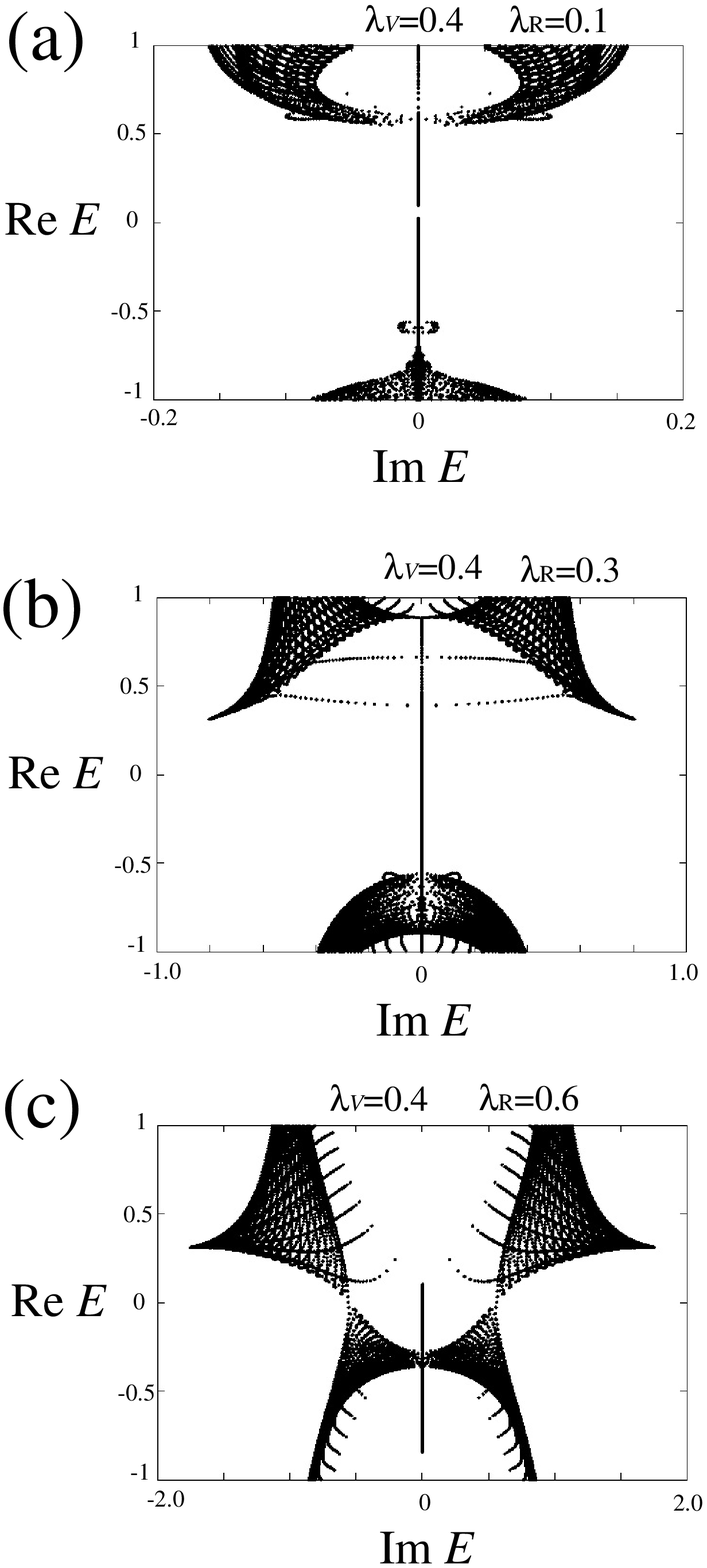}
\caption{The real part vs imaginary part of the energy bands
of non-Hermitian Kane-Mele model (\ref{so32_Hamiltonian}) with zigzag
  edges along the $x$-direction.
We plot the results for $t=1.0$, $\lambda_{\rm SO}=0.06$, 
$\lambda_V=0.4$, and various values of the non-Hermitian parameter 
$\lambda_{{\rm R}}$.
The data (b) indicate that the energy of the gapless edge
  states [Fig.\ref{fig:so32_7} (b)] is real.
On the other hand, the bulk state supports the imaginary part of the energy.}
\label{fig:so32_8}
\end{center}
\end{figure}

Next consider the region $\lambda_V> 3\sqrt{3}\lambda_{{\rm SO}}$.
In the presence of weak non-Hermiticity $\lambda_{\rm R}$,
 gapless edge modes do not appear [Figs.\ref{fig:so32_7}(a) and \ref{fig:so32_8}(a)].
This is because the bulk gap has not yet closed and the system remains a 
topologically trivial phase with $C_{\rm TRI}=0$.
If we further increase $\lambda_{\rm R}$, the bulk gap closes near
$a k_x\sim 2\pi/3$ and $4\pi/3$ at a critical value of $\lambda_{\rm
R}$, then the bulk gap opens again [Figs. \ref{fig:so32_7}(b) and
\ref{fig:so32_8}(b).] 
In this region of $\lambda_{\rm R}$,
gapless edge modes appear which are topologically protected 
by the time-reversal invariant Chern number.
For sufficiently large $\lambda_{\rm R}$, the bulk gap closes once again 
at $a k_x\sim \pi$, and the edge modes disappear as shown  
in Figs. \ref{fig:so32_7}(c) and \ref{fig:so32_8}(c).

\section{Summary and discussion
}
\label{sec:summary}

In this work, we investigated edge modes and their topological stability in non-Hermitian models. 
We analyzed three types of models: $SU(1,1)$ lattice model realized on graphene with pure imaginary sublattice potential,  $SO(3,2)$ Luttinger model on 2D square lattice, and $SO(3,2)$ Kane-Mele model with asymmetric hopping integrals on graphene.   
The energy spectra of such non-Hermitian models
generally contain complex eigenvalues. 
In this paper, we focused on the real parts of the edge bands and 
characterized them by using topological arguments. 
(The imaginary part of eigenvalues brings decay of wavefunction with time.)  
For the $SU(1,1)$ lattice model, with numerical calculations, we found that the edge states with Re$E=0$ are robust against small non-Hermitian perturbation.  We gave topological arguments for the robustness of edge state.
 Meanwhile,  
the $SO(3,2)$ Luttinger and the $SO(3,2)$ Kane-Mele models are time-reversal invariant non-Hermitian models with $\Theta^2=+1$. 
The generalized Kramers theorem suggests the existence of helical edge modes in the models. 
The numerical calculations indeed confirmed the existence of helical edge modes and robustness of them under the small non-Hermitian perturbations.  
We introduced time-reversal invariant Chern number inherent to non-Hermitian models, and gave topological arguments about the stability of helical edge modes. 

In this paper, we adopted non-Hermitian models whose Hermitian counterparts are typical topological insulators in 2D.   
There are many types of topological insulators, such as topological superconductors, 3D topological insulators, etc.  
It would be interesting to consider non-Hermitian generalizations of
these various topological insulators,
where gapless edge modes could also appear.
The topological arguments for non-Hermitian systems
presented in this paper would constitute the first step
to introduce topological invariants characterizing gapless
edge modes of other non-Hermitian models.

\section*{Note added} 
After the completion of this work, we noticed Ref.\cite{HuHughes} in which the $SU(1,1)$ and $SO(3,2)$ models are 
discussed in the context of topological insulators.  
Their numerical calculations of the lattice version of the $SU(1,1)$ coincide with our results.   
In Ref.\cite{HuHughes}, the authors concluded that the appearance of the complex eigenvalues 
is an indication of the non-existence of the topological insulator phase in non-Hermitian models. 
In the present paper, we focused on the real part of the complex eigenvalues and explored the robustness of the gapless edge modes on the basis of topological stability arguments.  
The ``topological phase'' in the present paper is referred to the phase in which real part of edge modes is stable under small non-Hermitian perturbations.

\begin{acknowledgments}

We acknowledge ISSP visiting program that enabled the collaboration. 
K.E.  was supported in part by Global COE Program
``the Physical Sciences Frontier,'' MEXT, Japan. 
This work was supported in part by a Grant-in Aid for Scientific
Research from MEXT of Japan, 
Grants No.22103005 (Innovative Areas ``Topological Quantum Phenomena''),
No.22540383, and No.23740212. 

\end{acknowledgments}

\appendix

\section{Imaginary Rashba interactions
and imaginary $SU(2)$ gauge potentials}
\label{sec:Appendix_SU2}

Let us first consider continuum Hamiltonians with {\it imaginary} Rashba
couplings $i \lambda$ ($\lambda$: real constants),
\be
H&=&\frac{1}{2m}(p_x^2+p_y^2) + i \lambda (p_x\sigma_y-p_y\sigma_x)\nonumber\\
 &=&\frac{1}{2m} (p_x+i\theta\sigma_y)^2+\frac{1}{2m} (p_y-i\theta\sigma_x)^2
+m\lambda^2,
\label{continuum_Hamiltonian}
\ee
where $\theta\equiv m\lambda$.
From (\ref{continuum_Hamiltonian}), we find that
the {\it imaginary} Rashba interaction 
can be regarded as an {\it imaginary} $SU(2)$ gauge potential:
\be
(\widetilde{A}_x,\widetilde{A}_y)
= \left(-i \theta \sigma_y,i \theta \sigma_x\right).
\label{gauge_continuum}
\ee
These {\it imaginary} gauge potential acquires {\it scale transformations}
$\widetilde{U_x}$ and $\widetilde{U_y}$ when we proceed by  
a unit length along the $x$ and $y$ directions, respectively:
\be
\widetilde{U}_x=e^{\theta \sigma_y},\quad 
\widetilde{U}_y=e^{-\theta \sigma_x}.
\ee

Now we perform scale transformations
$H'=\widetilde{U}(x,y)^{-1} H \widetilde{U}(x,y)$ with
\be
\widetilde{U}(x,y)=e^{\theta\sigma_y x} e^{-\theta\sigma_x y}.
\ee
For the first term of the Hamiltonian (\ref{continuum_Hamiltonian}),
we have
\be
\widetilde{U}(x,y)^{-1} (p_x+i\theta\sigma_y)^2 \widetilde{U}(x,y)
&=& e^{\theta\sigma_x y} e^{-\theta\sigma_y x} 
\left(-i \frac{\partial}{\partial x} +i\theta\sigma_y\right)^2
e^{\theta\sigma_y x} e^{-\theta\sigma_x y} \nonumber\\
&=& -e^{\theta\sigma_x y}\frac{\partial^2}{\partial x^2} e^{-\theta\sigma_x y}
= p_x^2.
\ee
For the second term of the Hamiltonian (\ref{continuum_Hamiltonian}),
we have
\be
&&\widetilde{U}(x,y)^{-1} (p_y-i\theta\sigma_x)^2  \widetilde{U}(x,y)
\nonumber\\
&&= e^{\theta\sigma_x y} e^{-\theta\sigma_y x} 
 (p_y-i\theta\sigma_x)^2 
e^{\theta\sigma_y x} e^{-\theta\sigma_x y}\nonumber\\
&&= e^{-2i\theta^2 x y \sigma_z} e^{-\theta\sigma_y x}  e^{\theta\sigma_x y}
 \left(-i\frac{\partial}{\partial y}-i\theta\sigma_x\right)^2 
 e^{-\theta\sigma_x y} e^{\theta\sigma_y x}  e^{2i\theta^2 x y \sigma_z}
\nonumber\\
&&=  - e^{-2i\theta^2 x y \sigma_z} e^{-\theta\sigma_y x} 
\frac{\partial^2}{\partial y^2} 
 e^{\theta\sigma_y x}  e^{2i\theta^2 x y \sigma_z} \nonumber\\
&&=   - e^{-2i\theta^2 x y \sigma_z}  
\frac{\partial^2}{\partial y^2}   e^{2i\theta^2 x y \sigma_z} \nonumber\\
&&= -\left(\frac{\partial}{\partial y} +2 i \theta^2 \sigma_z x \right)^2
=\left(p_y + 2 \theta^2 \sigma_z x \right)^2,
\ee
where we used the Baker-Campbell-Hausdorff formula 
up to the leading order of $\theta$.
Therefore, we have
\be
H'=\frac{1}{2m} p_x^2+\frac{1}{2m} \left(p_y + 2 \theta^2 \sigma_z x \right)^2
+m\lambda^2.
\label{continuum_Hamiltonian_2}
\ee

The Hamiltonian (\ref{continuum_Hamiltonian_2}) is Hermitian and
coincides with the one with {\it real} Rashba coupling $\lambda$
(under the transformation $\sigma_z\to -\sigma_z$).
We recall that Rashba interactions 
 with {\it real} couplings $\lambda$ can be regarded as $SU(2)$ 
gauge potentials\cite{NH_spin},
\be
(A_x,A_y)=\left(- \theta \sigma_y, \theta \sigma_x\right),
\ee
which give field strengths,
\be
F_{xy}&=&\partial_x A_y - \partial_y A_x -i[A_x,A_y]\nonumber\\
&=&i \theta^2 [ \sigma_y, \sigma_x]
= 2 \theta^2 \sigma_z.
\label{field_continuum}
\ee
Actually, (\ref{continuum_Hamiltonian_2}) with $\sigma_z\to -\sigma_z$
gives field strengths $F_{xy}= 2 \theta^2 \sigma_z$ (\ref{field_continuum}).

Although non-Hermitian Hamiltonian $H$ is transformed 
into Hermitian Hamiltonian $H'$
by the scale transformations $\widetilde{U}(x,y)$,
non-Hermiticity affects the boundary conditions.
Suppose that right eigenfunctions $\psi^R(x,y)$ 
and left eigenfunctions $\psi^L(x,y)$ of $H$ have periodic 
boundary conditions both in $x$ and the $y$ directions:
\be
\psi^R(L_x,y)=\psi^R(0,y),\quad \psi^L(L_x,y)=\psi^L(0,y), \nonumber\\
\psi^R(x,L_y)=\psi^R(x,0),\quad \psi^L(x,L_y)=\psi^L(x,0).
\ee
From the boundary condition in the $x$ direction,
eigenfunctions $\psi'(x,y)=\widetilde{U}(x,y)^{-1}\psi^R(x,y)$ of $H'$
are found to satisfy 
\be
 \psi'(L_x,y)
&=&\widetilde{U}(L_x,y)^{-1}\psi^R(L_x,y)\nonumber\\
&=&\widetilde{U}(L_x,y)^{-1}\widetilde{U}(0,y)
\widetilde{U}(0,y)^{-1}\psi^R(0,y)\nonumber\\
&=&  e^{\theta\sigma_x y} e^{-\theta\sigma_y L_x} e^{-\theta\sigma_x y} \psi'(0,y)\nonumber\\
&=&  e^{-\theta\sigma_y L_x} e^{-2i\theta^2\sigma_z L_x y} \psi'(0,y),
\ee
and
\be
 \psi'(L_x,y)^{\dag}
=\psi'(0,y)^{\dag} e^{2i\theta^2\sigma_z L_x y}  e^{\theta\sigma_y L_x}.
\ee
Similarly, the boundary condition in the $y$ direction gives,
\be
 \psi'(x,L_y)= e^{\theta\sigma_x L_y} \psi'(x,0),
\quad  \psi'(x,L_y)^{\dag}= \psi'(x,0)^{\dag} e^{-\theta\sigma_x L_y} .
\ee

\begin{figure}
 \begin{center}
  \includegraphics[width=7.0cm,clip]{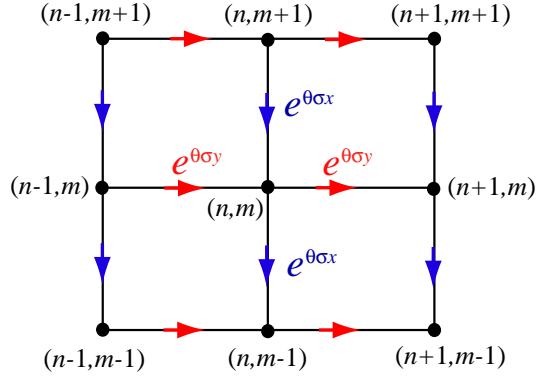}
\caption{\label{fig:su2_square}
Tight-binding model on the square lattice with {\it imaginary}
$SU(2)$ gauge fields.
}
\end{center}
\end{figure} 
Let us now generalize the above arguments to lattice systems.
For simplicity, we consider a tight-binding model on the square lattice
with the nearest-neighbor hopping integrals with {\it imaginary}
$SU(2)$ gauge fields:
\be
H&=&\sum_{(n,m)} \left[c_{(n+1,m)}^{\dag} e^{\theta\sigma_y} c_{(n,m)}
+c_{(n-1,m)}^{\dag} e^{-\theta\sigma_y} c_{(n,m)} \right. \nonumber\\
&{}&~~~~~~ \left. +c_{(n,m+1)}^{\dag} e^{-\theta\sigma_x} c_{(n,m)}
+c_{(n,m-1)}^{\dag} e^{\theta\sigma_x} c_{(n,m)} \right]  \nonumber\\
&+& \sum_{(n,m)} \left[c_{(n,m)}^{\dag} e^{-\theta\sigma_y} c_{(n+1,m)}
+c_{(n,m)}^{\dag} e^{\theta\sigma_y} c_{(n-1,m)} \right.\nonumber\\
&{}&~~~~~~\left. +c_{(n,m)}^{\dag} e^{\theta\sigma_x} c_{(n,m+1)}
+c_{(n,m)}^{\dag} e^{-\theta\sigma_x} c_{(n,m-1)}\right],
\label{SU2_original}
\ee
which is illustrated in Fig.\ref{fig:su2_square}.
By scale transformations,
\be
c_{(n,m)}=e^{\theta \sigma_y n} e^{-\theta \sigma_x m} \widetilde{c}_{(n,m)},
\quad
c_{(n,m)}^{\dag}= \widetilde{c}_{(n,m)}^{\dag}
e^{\theta \sigma_x m} e^{-\theta \sigma_y n},
\ee
the Hamiltonian (\ref{SU2_original}) is transformed into
\be
H' &=&\sum_{(n,m)} \left[\widetilde{c}_{(n+1,m)}^{\dag} \widetilde{c}_{(n,m)}
+\widetilde{c}_{(n-1,m)}^{\dag} \widetilde{c}_{(n,m)} \right. \nonumber\\
&{}&~~~~~~ \left. +\widetilde{c}_{(n,m+1)}^{\dag} e^{-2 i\theta^2 \sigma_z n} 
\widetilde{c}_{(n,m)}
+\widetilde{c}_{(n,m-1)}^{\dag} e^{2 i\theta^2 \sigma_z n} 
\widetilde{c}_{(n,m)} \right]  \nonumber\\
&+& \sum_{(n,m)} \left[\widetilde{c}_{(n,m)}^{\dag} \widetilde{c}_{(n+1,m)}
+\widetilde{c}_{(n,m)}^{\dag} \widetilde{c}_{(n-1,m)} \right.\nonumber\\
&{}&~~~~~~\left. +\widetilde{c}_{(n,m)}^{\dag} 
e^{2 i\theta^2 \sigma_z n} \widetilde{c}_{(n,m+1)}
+\widetilde{c}_{(n,m)}^{\dag} e^{-2 i\theta^2 \sigma_z n} 
\widetilde{c}_{(n,m-1)}\right],
\label{Hermitian_square}
\ee
where we used the Baker-Campbell-Hausdorff formula 
up to the leading order of $\theta$.
The Hamiltonian (\ref{Hermitian_square}) is the tight-binding model
with $SU(2)$ gauge fields,
which gives an observable flux $2 \theta^2 \sigma_z$ per plaquette
(under a transformation $\sigma_z\to -\sigma_z$).

Let us impose the periodic boundary condition
on the original Hamiltonian $H$ (\ref{SU2_original}):
\be
&&c_{(L_x,m)}=c_{(0,m)}, \quad c_{(L_x,m)}^{\dag}=c_{(0,m)}^{\dag}, \nonumber\\
&&c_{(n,L_y)}=c_{(n,0)}, \quad c_{(n,L_y)}^{\dag}=c_{(n,0)}^{\dag}.
\ee
Then, non-Hermiticity appears as boundary conditions
for the Hamiltonian $H'$ (\ref{Hermitian_square}):
\be
&&\widetilde{c}_{(L_x,m)}
=e^{-\theta \sigma_y L_x} e^{-2i\theta^2\sigma_z L_x m}
\widetilde{c}_{(0,m)}, \quad
\widetilde{c}_{(L_x,m)}^{\dag}
=\widetilde{c}_{(0,m)}^{\dag} e^{2i\theta^2\sigma_z L_x m}
e^{\theta \sigma_y L_x},\nonumber\\
&&\widetilde{c}_{(n,L_y)}
=e^{\theta \sigma_x L_y}\widetilde{c}_{(n,0)}, \quad
\widetilde{c}_{(n,L_y)}^{\dag}
=\widetilde{c}_{(n,0)}^{\dag} e^{-\theta \sigma_x L_y}.
\ee

\end{document}